\documentclass[aoas]{imsart}

\RequirePackage{amsthm,amsmath,amsfonts,amssymb}
\RequirePackage[authoryear]{natbib}
\RequirePackage[colorlinks,citecolor=blue,urlcolor=blue]{hyperref}
\RequirePackage{graphicx}
\usepackage{bm}
\usepackage{threeparttable}
\usepackage{booktabs}
\usepackage{paralist}
\usepackage{comment}
\usepackage[dvipsnames]{xcolor}

\def\bSig\mathbf{\Sigma}

\def\bSig\mathbf{\Sigma}

\newcommand\mc{\mathcal}
\newcommand{\E}{\mathbb{E}}
\newcommand\ve{\varepsilon}

\newcommand\wh{\widehat}

\newcommand{\bfX}{\bm{X}}

\startlocaldefs
\theoremstyle{plain}

\newtheorem{theorem}{Theorem}
\newtheorem{proposition}{Proposition}

\newtheorem{assumption}{Assumption}

\theoremstyle{remark}

\newtheorem{remark}{Remark}

\endlocaldefs

\begin{document}

\begin{frontmatter}
\title{Multiply robust estimation for causal survival analysis with treatment noncompliance}
\runtitle{Causal survival analysis with noncompliance}

\begin{aug}
\author[A]{\fnms{Chao}~\snm{Cheng}\ead[label=e1]{chaoc@wustl.edu}},
\author[B]{\fnms{Bo}~\snm{Liu}\ead[label=e2]{bo.liu1997@duke.edu}},
\author[C,D]{\fnms{Lisa}~\snm{Wruck}\ead[label=e3]{lisa.wruck@duke.edu}}
\author[B]{\fnms{Fan}~\snm{Li}\ead[label=e4]{fl35@duke.edu}}
\and
\author[E]{\fnms{Fan}~\snm{Li}\ead[label=e5]{fan.f.li@yale.edu}}
\address[A]{Department of Statistics and Data Science, Washington University in St. Louis\printead[presep={,\ }]{e1}}
\address[B]{Department of Statistical Science, Duke University\printead[presep={,\ }]{e2,e4}}
\address[C]{Department of Biostatistics and Bioinformatics, Duke University\printead[presep={,\ }]{e3}}
\address[D]{Duke Clinical Research Institute}
\address[E]{Department of Biostatistics, Yale University\printead[presep={,\ }]{e5}}

\end{aug}

\begin{abstract}
Comparative effectiveness research frequently addresses a time-to-event outcome and can require unique considerations in the presence of treatment noncompliance. Motivated by the challenges in addressing noncompliance in the ADAPTABLE pragmatic clinical trial, we develop a multiply robust estimator to estimate the principal survival causal effects under the principal ignorability and monotonicity. The multiply robust estimator is consistent even if one, and sometimes two, of the required models are misspecified. We apply the multiply robust method in the ADAPTABLE trial to evaluate the effect of low- versus high-dose aspirin assignment on patients’ death and hospitalization from cardiovascular diseases. We find that, comparing to low-dose assignment, assignment to the high-dose leads to differential effects among always high-dose takers, compliers, and always low-dose takers. Such treatment effect heterogeneity contributes to the null intention-to-treatment effect. We further perform a formal sensitivity analysis for investigating the robustness of our causal conclusions under violation of two identification assumptions specific to noncompliance. 
\end{abstract}


\end{frontmatter}

\section{Introduction}
\label{sec:intro}
\subsection{ADAPTABLE pragmatic trial and motivating question}\label{sec:intro1}

ADAPTABLE (\textit{Aspirin Dosing: A Patient-Centric Trial Assessing Benefits and Long-Term Effectiveness}) is an open-label, pragmatic, randomized trial to study the effectivenss of two strategies of aspirin dosing---325 mg versus 81 mg per day---for lowering risk on death and hospitalization among patients with existing cardiovascular diseases (\citealp{jones2021comparative}). A total of 15,076 participants were enrolled in study, where 7,540 were randomized to the 81-mg group (low dose) and 7,536 were randomized to the 325-mg group (high dose). The primary outcome is a composite of death from any cause and hospitalization for stroke or myocardial infarction, assessed by time to first event. Standard intention-to-treat (ITT) analysis suggested no statistically significant difference between the two dosing strategies in reducing patients' risk on death and cardiovascular events (\citealp{jones2021comparative}).

{As a pragmatic trial, the ADAPTABLE includes a substantial proportion of participants who did not adhere to their assigned aspirin dosage after randomization.} We are interested in evaluating the causal effects of aspirin dosing assignment among subgroups of patients according to their treatment compliance status and separating the treatment efficay from the direct effect of treatment assignment. We consider the principal stratification framework (\citealp{FrangakisRubin02}) to subset patients into subgroups, including always high-dose takers who would stick to the 325-mg aspirin dosage regardless of their assignment, always low-dose takers who would stick to low-dose regardless of their assignment, and compliers who would comply with their assignment. The causal effect within each subgroup is referred to as the principal causal effect, which is a meaningful measure that can offer insights into whether one specific aspirin  dosage benefits patients in each subgroup. { In the noncompliance literature, the \textit{exclusion restriction} (ER) is frequently used to identify the principal causal effect, which rules out any direct effect of treatment assignment on the outcome other than through the treatment actually received (\citealp{Hirano00,ding2017principal,jiang2020multiply}). However, the ER assumption may be questionable in open-label studies such as ADAPTABLE \citep{liu2024principal}. In pragmatic trials such as ADAPTABLE, many participants had comorbidities, and their healthcare providers---aware of the assigned aspirin dosage---may have adjusted monitoring intensity or care plans based on perceived risk associated with assignment. Such uncontrolled, and often unmeasured factors in pragmatic, routine healthcare settings may have introduced direct effects of treatment assignment on patient outcomes.}

In ADAPTABLE, several complications arise for estimating the principal causal effects. {First, addressing noncompliance in open-label studies requires identification assumptions beyond ER, as we also aim to assess potential direct effects of treatment assignment.} Second, identification assumptions are unverifiable from the observed data, and tailored sensitivity methods are useful to provide a context for interpreting the study results. Third, the primary outcome is time-to-event and requires unique handling as it is only partially observed due to right censoring from study dropout or end of follow-up. Despite randomization, patients' censoring status may still depend on their baseline characteristics \citep{robins2000correcting}, which must be accounted for. Last but not the least, the validity of existing methods for addressing noncompliance often heavily relies on the correct model specification. For example, to estimate the principal causal effects with a time-to-event outcome, one may need to specify multiple working models, e.g., those for the treatment assignment process, compliance status, censoring process, or the time-to-event process. Whereas the treatment assignment is known under randomization, all other models must be estimated from data and their misspecifications may lead to biased causal effect estimates \citep{jo2009use}. More robust survival methods that can protect from working model misspecifications are of interest for applications to the ADAPTABLE study.

\subsection{Related literature and our contributions}\label{sec:intro2}

Several methods have been proposed to identify principal causal effects in the presence of noncompliance. Under \textit{monotonicity} and the ER assumption, the instrumental variable approach was first used to identify the complier average causal effect (\citealp{imbens1994identification,baker1994paired,Angrist96}), which was later generalized to a principal stratification framework to deal with more general post-treatment confounding problems (\citealp{FrangakisRubin02}). Despite the extensions of \cite{Angrist96} and \cite{FrangakisRubin02}'s work in a variety of directions, there are only a few methods that are geared toward studying noncompliance with a time-to-event outcome. \cite{baker1998analysis} proposed a likelihood-based approach for the complier average causal effect on a hazard scale without covariates. Later, several authors considered semiparametric mixture models to estimate complier treatment effect defined on survival probabilities; for example, the proportional hazard model used in \cite{loeys2003causal} and \cite{cuzick2007estimating} and semiparametric transformation models used in \cite{yu2015semiparametric}. A couple of nonparametric approaches have also been developed to address noncompliance with time-to-event outcomes; for example, the Kaplan–Meier estimator in \cite{frangakis1999addressing}, the empirical likelihood approach in \cite{nie2011inference}, and the nonparametric estimator for complier quantile causal effect in \cite{wei2021estimation}. {Recently, \cite{liu2024principal} developed a Bayesian survival mixture model to empirically identify principal causal effects with time-to-event outcomes.}

In the majority of these previous work, ER serves as a key assumption to identify the causal effect (see Web Table 1 for a survey of these existing developments), and assumes that the treatment assignment exerts no direct effect on patients who would receive the same treatment regardless of assignment. {However, as previously discussed, this assumption may not be easily justified in open-label trials like ADAPTABLE. As an alternative to ER, we consider the principal ignorability assumption to point identify the principal causal effect  (\citealp{jo2009use,ding2017principal}). Principal ignorability assumes that the observed pre-treatment covariates are sufficient to controlling for confounding due to the patients' compliance status without excluding any direct effect of assignment. This allows us to estimate, in addition to the complier average causal effect, the causal effect of assignment among the always high-dose takers and always low-dose takers. The ADAPTABLE trial collects a rich set of baseline covariates (Web Table 2), including detailed patient demographic and medical characteristics. This comprehensive baseline information in ADAPTABLE provides a strong foundation for assessing causal effects under the principal ignorability assumption.} 

{Despite recent advancement of causal inference methods under principal ignorability \citep{ding2017principal,jiang2020multiply}, little development has been made to address right-censored survival outcomes. In this paper, we contribute a new multiply robust estimator for the principal survival causal effect, and implement this new estimator to estimate the principal causal estimands in ADAPTABLE.} {Although this work is motivated by a randomized trial, the proposed estimator is also applicable to treatment noncompliance in quasi-experiments \citep{angelucci2012impact,angelucci2006estimating}, where the treatment assignment process is not randomized but is assumed conditionally ignorable given observed covariates.} The core components of our multiply robust estimator include working models for the treatment propensity score, principal score, censoring process and survival time of interest. We show that it is multiply robust in the sense it is consistent to the principal causal effects even if one, and sometimes two, of the working models are misspecified. In this sense, our estimator provides stronger protection against model misspecification than a full likelihood approach. Finally, because the validity of our estimator relies on the principal ignorability and monotonicity assumptions, we develop a sensitivity analysis framework to assess the impact of their potential violations, and illustrate its application using the ADAPTABLE study.

\section{Notation, estimands, and structural assumptions}\label{sec:2}

Consider a comparative effectiveness study with $n$ patients for comparing two treatments (high versus low aspirin dosage). For each patient $i=1,\dots,n$, we record a vector of pre-treatment covariates $\bm X_i$. Let $Z_i \in \{0,1\}$ be the treatment assignment for patient $i$. In the context of ADAPTABLE, $Z_i = 1$ if the patient is assigned to the high-dose group and $Z_i = 0$ if assigned to the low-dose group. Let $S_i \in \{0,1\}$ be the actual treatment that patient $i$ received, with $S_i =1$ if the patient received the high aspirin dose and $S_i = 0$ if received the low aspirin dose. Each patient is associated with a failure time $T_i$ that is incompletely observed due to right-censoring at time $C_i$. Therefore, we only have the observed failure time $U_i = \min(T_i,C_i)$ and a censoring indicator $\delta_i=\mathbb{I}(T_i\leq C_i)$, where $\mathbb{I}(\cdot)$ is the indicator function. The observed data consists of $n$ independent and identically distributed copies of the quintuple ${\mc O}_i=\{\bfX_i,Z_i,S_i,U_i,\delta_i\}$. 

We adopt the potential outcomes framework to define the causal estimands, and assume the Stable Unit Treatment Value Assumption (SUTVA) such that the assignment is defined unambiguously and there is no patient-level interference. 
For patient $i$, let $S_i(z)$ be the potential treatment receipt if the patient is assigned to treatment $z \in \{0,1\}$. Therefore, $S_i(z) = 1$ indicates that the patient $i$ would receive the high dose if assigned to treatment $z$ and $S_i(z) = 0$ otherwise. Under SUTVA, we can connect the observed treatment receipt $S_i$ and potential treatment receipt $S_i(z)$ by recognizing $S_i=Z_iS_i(1)+(1-Z_i)S_i(0)$. Similarly, we define the potential failure time and the potential censoring time for patient $i$ under assignment $z$ by $T_i(z)$ and $C_i(z)$. Under SUTVA, we also have $T_i=Z_iT_i(1)+(1-Z_i)T_i(0)$ and $C_i=Z_iC_i(1)+(1-Z_i)C_i(0)$.

Under the principal stratification framework \citep{FrangakisRubin02}, we use the joint potential values of the treatment receipt, $G_i=\left(S_i(1),S_i(0)\right) \in \{0,1\}^{\otimes 2}$, to define subgroups, or alternatively referred to as principal strata. In the context of ADAPTABLE, we refer to the four possible principal strata, $G_i= \{(1,1),(1,0),(0,1),(0,0)\}$, as the \textit{always high-dose takers} who take the 325-mg aspirin dosage regardless of randomized assignment, the \textit{compliers} who comply with the randomized aspirin dosage, the \textit{defiers} who take the the opposite aspirin dosage to their randomized assignment, and the \textit{always low-dose takers} who take the 81-mg aspirin dosage regardless of randomized assignment. Hereafter, we abbreviate the four principal strata as $G\in\{a,c,d,n\}$. A central property of the principal strata is that it is unaffected by assignment and therefore can be considered as a pre-treatment covariate, conditional on which the subgroup causal effects are well-defined. Within each strata, our interest lies in the principal survival causal effect (PSCE): 
\begin{equation}\label{eq:psce}
\Delta_g(u)=\mathcal{S}_{1,g}(u)-\mathcal{S}_{0,g}(u)=\mathbb{P}(T(1)\ge u|G=g)-\mathbb{P}(T(0)\ge u|G=g), 
\end{equation}
for $g\in\{n,c,d,a\}$, where $u$ is a pre-specified time point at which the counterfactual survival function is evaluated. In words, the PSCE estimand quantifies the survival benefit at time $u$ when the subpopulation with the compliance pattern $g$ is placed under the high-dose group versus when placed under the low-dose group. In particular, $\Delta_{c}$ has been referred to as the complier average causal effect (CACE) in survival probability \citep{yu2015semiparametric}. In a similar fashion, if we consider the low-dose group as the ``usual care'' group, we can also refer to $\Delta_{a}$ and $\Delta_{n}$ as the always-taker average causal effect (AACE) and the never-taker average causal effect (NACE), which characterize the the direct effect of treatment assignment that could possibly be attributed to mechanisms that are unmeasured in the study. With a survival outcome, identification of PSCE requires the following assumptions. 

\begin{assumption}\label{assum:2}
(Unconfoundedness and overlap) $\{T(1),T(0),S(1),S(0)\}\perp Z|\bfX$, where the symbol ``$\perp$'' denotes independence. Furthermore, the propensity score $\pi(\bfX)=\mathbb{P}(Z=1|\bfX)$ is strictly bounded away from $0$ and $1$.
\end{assumption} 
{Assumption \ref{assum:2} assumes that the treatment assignment is ignorable conditional on observed baseline covariates. In a completely randomized trial such as ADAPTABLE, a stronger assumption, $\{T(1),T(0),S(1),S(0),\bfX\}\perp Z$, holds as a result of the study design. In quasi-experiments, Assumption \ref{assum:2} is plausible when a rich set of baseline covariates has been collected to approximate the unknown assignment mechanism. In what follows, we maintain this more general ignorability assumption to unify developments for both randomized trials and quasi-experiments.}  

\begin{assumption}\label{assum:3}
(Monotonicity) $S_i(1)\ge S_i(0)$ for all $i$.
\end{assumption}
The {monotonicity} assumption excludes defiers. In the context of ADAPTABLE, the defiers, if exist, would consist of patients who take the 81-mg aspirin if assigned to the high-dose group, but take the 325-mg aspirin if assigned to the low-dose group.
This type of patients is considered unlikely in ADAPTABLE, though it might be applicable in other settings \citep{tong2025semiparametric}. Of note, monotonicity cannot be empirically verifiable based on the observed data alone, and we will consider sensitivity analysis strategies to assess assumed departure from this assumption in Section \ref{sec:SA-Mono}.

Under Assumption \ref{assum:3}, each individual's principal strata membership is only partially identified. That is, the observed cells with $(Z=1,S=0)$ and $(Z=0,S=1)$ include only the always low-dose takers and the always high-dose takers, respectively. The observed cells with $(Z=0,S=0)$ and $(Z=1,S=1)$, however, are a mixture of patients in two principal strata---the $(Z=0,S=0)$ cell consists of the always low-dose takers and the compliers, while the $(Z=1,S=1)$ cell consists of the always high-dose takers and the compliers. The following assumption thus is central to disentangle the latent membership within the $(Z=0,S=0)$ and $(Z=1,S=1)$ cells.
\begin{assumption}\label{assum:4}
(Prinicipal ignorability) For all $u\geq 0$, we have $\mathbb{P}\left(T(1)\ge u|G=a,\bfX\right)=\mathbb{P}\left(T(1)\ge u|G=c,\bfX\right)$ and $\mathbb{P}\left(T(0)\ge u|G=c,\bfX\right)=\mathbb{P}\left(T(0)\ge u|G=n,\bfX\right)$.
\end{assumption} 
Assumption \ref{assum:4} assumes that, conditional on pre-treatment covariates, the survival functions for $T(1)$ become exchangeable between the always high-dose takers and the compliers conditional on covariates, and likewise, the survival functions for $T(0)$ become exchangeable between the always low-dose takers and the compliers \citep{jo2009use,ding2017principal}. This requires that a sufficient set of covariates, $\bfX$, have been collected to capture the confounding between noncompliance status and the survival outcome. As a pragmatic trial leveraging routinely-collected data, ADAPTABLE included patients' demographics, smoking status, medical history and aspirin use prior to randomization, which are critical in explaining the compliance patterns. However, since the principal ignorability is also not empirically verifiable based on the observed data itself, we will explore sensitivity analysis in Section \ref{sec:sa_pi}. 

Under Assumptions \ref{assum:2}--\ref{assum:3}, Assumption \ref{assum:4} implies that 
\begin{align}
    \mathbb{P}(T(1)\ge u|G=g,Z=1,S=1,\bfX)=\mathbb{P}(T\ge u|Z=1,S=1,\bfX), \text{ for } g\in\{a,c\}, \label{eq:pi1}\\
    \mathbb{P}(T(0)\ge u|G=g,Z=0,S=0,\bfX)=\mathbb{P}(T\ge u|Z=0,S=0,\bfX), \text{ for } g\in\{n,c\}, \label{eq:pi2}
\end{align} 
for all $u\geq 0$. These two equations allow us to identify a set of conditional survival functions for the potential survival times, $T(1)$ and $T(0)$. For example, equation \eqref{eq:pi1} suggests that, within the observed subpopulation who were assigned to the high-dose group and also received the high dose, the compliance strata variable does not further affect $T(1)$ given pre-treatment covariates. Therefore the conditional survival function given the compliance status $G$, $\mathbb{P}(T(1)\ge u|G=g,Z=1,S=1,\bfX)$, simplifies to a function of the observed data only, $\mathbb{P}(T\ge u|Z=1,S=1,\bfX)$. 

Finally, noting that the survival outcome may be partially observable due to right censoring, we adopt the following conditionally independent censoring assumption: 
\begin{assumption}\label{assum:5}
(Conditionally independent censoring) $T(z)\perp C(z)|\{Z=z,S,\bm X\}$.
\end{assumption}
Assumption \ref{assum:5} stipulates that the potential censoring time does not provide any information about the potential failure time other than that the latter exceeds the censoring time conditional on pre-treatment covariates, and strata defined based on treatment assignment and the treatment receipt. Assumption \ref{assum:5} resembles the \textit{independent censoring} or \textit{coarsening at random} assumption in the standard survival analysis literature (\citealp{gill1997coarsening}).

\begin{remark}
\emph{(Alternatively identification assumption) In addition to Assumption \ref{assum:5}, an alternative censoring assumption can be used for identifying principal causal effects. Specifically, with non-censored outcomes, \cite{nguyen2024identification} introduced the latent missing-at-random (LMAR) assumption. When adapted to the context of right-censored survival outcomes, their assumption can be expressed as $T(z) \perp C(z) \mid \{Z = z, G, \bm X\}$. Compared to Assumption \ref{assum:5}, LMAR assumes that the censoring mechanism is conditionally independent of the event time within each latent principal stratum, rather than conditional on the observed treatment receipt status. Despite this difference, we can show that under Assumptions \ref{assum:2}--\ref{assum:3} and a stronger principal ignorability assumption (Assumption 5 in Web Appendix A.1), the LMAR assumption directly implies Assumption \ref{assum:5}.
The proof is provided in Web Appendix A.1. Compared to Assumption \ref{assum:4}, Assumption 5 in Web Appendix A.1 requires that the bivariate survival function of $\{T(z), C(z)\}$, rather than the marginal survival function of $T(z)$ alone, is conditionally exchangeable across principal strata. Due to this connection, all the following identification results continue to hold under Assumptions \ref{assum:2}, \ref{assum:3}, 5, plus the LMAR assumption.}
\end{remark}

\section{Estimating Principal Survival Causal Effects}\label{sec:3}
\subsection{Models specification for aspects of the data generating process}
\label{sec:3.1}

To estimate the defined PSCE estimands, there are four possible models that can be estimated from the observed data; each model represents a distinct aspect of the data generating process. In the context of ADAPTABLE, these models are:
\begin{itemize}
\item[(a)] $\pi_z(\bfX)=\mathbb{P}(Z=z|\bfX)$: the probability of assignment to the high-dose group ($z=1$) or the low-dose group ($z=0$), where $\pi_1(\bfX)$ is referred to as the propensity score. Also, we define $\pi_z=\E[\pi_z(\bfX)]=\mathbb{P}(Z=z)$ as the marginal allocation proportion. {In ADAPTABLE, we have $\pi_1(\bfX) = \pi_1 = 0.5$ by randomization, but more generally $\pi_z(\bfX)$ can be unknown and depend on $\bfX$ in quasi-experiments.}
\item[(b)] $e_g(\bfX)=\mathbb{P}(G=g|\bfX)$, for $g\in\{a,c,n\}$: the principal score of being in the always high-dose, complier, and always low-dose strata, respectively. Similarly, we write $e_g=\E[e_g(X)]=P(G=g)$ as the marginal strata proportion. 
\item[(c)] $\mc S^C_{zs}(u|\bfX)=\mathbb{P}(C\ge u|Z=z,S=s,\bfX)$, for $\{z,s\} \in \{0,1\}^{\otimes 2}$:  the conditional survival function of the censoring time given the assignment, treatment receipt and covariates.
\item[(d)] $\mc S_{zs}(u|\bfX)=\mathbb{P}(T\ge u|Z=z,S=s,\bfX)$, for $\{z,s\} \in \{0,1\}^{\otimes 2}$:  the conditional survival function of the outcome given the assignment, treatment receipt and covariates.
\end{itemize}

Logistic regression can be used to model the propensity score such that $\pi_1^{\text{par}}(\bfX;\bm\alpha) = 1/(1+e^{-\bm\alpha^T\bfX})$, where $\widehat{\bm\alpha}$ can obtained by maximum likelihood estimation. After obtaining $\widehat{\bm\alpha}$, we have $\widehat\pi_z(\bfX) = \{\pi_1^{\text{par}}(\bfX;\widehat{\bm\alpha})\}^z \{1-\pi_1^{\text{par}}(\bfX;\widehat{\bm\alpha})\}^{1-z}$. Although the true propensity score is known in randomized trials, one may preferably use the estimated propensity score to control for chance imbalance in covariates and to improve efficiency \citep{zeng2021propensity,li2022generalizing}. 

To estimate the principal scores, we utilize the following relationship \citep{ding2017principal}:
\begin{equation}\label{eq:bijection}
 e_{a}(\bfX)=p_{01}(\bfX),\quad e_{n}(\bfX)=p_{10}(\bfX),\quad e_c(\bfX)=p_{11}(\bfX)-p_{01}(\bfX).
\end{equation}
Here, $p_{zs}(\bfX)=\mathbb P(S=s|Z=z,\bfX)$ is the observed probability of receiving aspirin dosage $s\in\{0,1\}$ conditional on assignment to group $z \in \{0,1\}$ and covariates. Therefore, one can specify two logistic models such that $p_{z1}^{\text{par}}(\bfX;\bm\gamma_z)=1/(1+e^{-\bm\gamma_z^T\bfX})$ for $p_{z1}(\bfX)$ with $z\in\{0,1\}$, where $\widehat{\bm\gamma}_z$ is obtained by regressing $S$ on $\bfX$ based on the subset with $Z=z$. Then, for any $(z,s)\in\{0,1\}^{\otimes 2}$, we obtain $\widehat p_{zs}(\bfX) = \{p_{z1}^{\text{par}}(\bfX;\widehat{\bm\gamma}_z)\}^s\{1-p_{z1}^{\text{par}}(\bfX;\widehat{\bm\gamma}_z)\}^{1-s}$ as well as $\widehat e_g(\bfX)$ by \eqref{eq:bijection}.

The last two models correspond to regression for the observed censoring time and survival time across each observed cell $(Z,S) \in \{0,1\}^{\otimes 2}$. For each $(z,s)$, we consider the Cox proportional hazard model for the censoring process such that $\mc S^{C,\text{par}}_{zs}(u|\bfX;\bm\theta_{zs})=\exp\left(-\Lambda_{zs,0}^C(u)e^{-\widetilde{\bm\theta}_{zs}^T\bfX}\right)$, where $\bm\theta_{zs}=[\widetilde{\bm\theta}_{zs}^T,\Lambda_{zs,0}^C]^T$ contains all unknown parameters. The coefficients $\widetilde{\bm\theta}_{zs}$ can be estimated by maximum partial likelihood within the observed cell $(Z=z,S=s)$, and the cumulative baseline hazard ${\Lambda}_{zs,0}^C(u)$ can be estimated by the Breslow estimator \citep{breslow1972discussion}. A similar working proportional hazards model for the survival outcome can be considered as $\mc S_{zs}^{\text{par}}(u|\bfX;\bm\beta_{zs})=\exp\left(-\Lambda_{zs,0}(u)e^{-\widetilde{\bm\beta}_{zs}^T\bfX}\right)$ with unknown parameters $\bm\beta_{zs}=[\widetilde{\bm\beta}_{zs}^T,\Lambda_{zs,0}]^T$; the estimation of $\bm\beta_{zs}$ is analogous to that of $\bm\theta_{zs}$. We denote the estimated survival distribution of the censoring time and event time as $\widehat{\mc S}^{C}_{zs}(u|\bfX) = \mc S^{\text{par}}_{zs}(u|\bfX;\widehat{\bm\theta}_{zs})$ and $\widehat{\mc S}_{zs}(u|\bfX) = \mc S_{zs}^{\text{par}}(u|\bfX;\widehat{\bm\beta}_{zs})$, respectively. 

Throughout, we use $\mathcal M_\pi,\mc M_e,\mc M_C,\mc M_T$ to denote the models for $\pi_z(\bfX)$, $p_{zs}(\bfX)$, $\mc S^C_{zs}(u|\bfX)$, and  $\mc S_{zs}(u|\bfX)$, respectively. To facilitate presentation, we define $p_{zs}=\mathbb{E}[p_{zs}(\bfX)]$, which is a key component for estimating the PSCE \eqref{eq:psce}. Following \cite{jiang2020multiply}, we consider a doubly robust estimator for $p_{zs}$:\begingroup\makeatletter\def\f@size{10}\check@mathfonts
\begin{equation}\label{eq:p0_p1_dr}
\wh p_{zs}=\mathbb{P}_n\left[\frac{\mathbb{I}(Z=z)(\mathbb{I}(S=s)-p_{zs}(\bfX))}{\pi_z(\bfX)}+p_{zs}(\bfX)\right]
\end{equation}\endgroup
which leads to unbiased estimation under the union model $\mc M_{\pi}\cup\mc M_{e}$; i.e., it is consistent if either $\mc M_{\pi}$ or $\mc M_{e}$ is correctly specified, but not necessary both. Throughout, we use the union notation, ``$\cup$", to denote correct specification of at least one model. The estimators for the marginal proportion of each prinicpal stratum are $\widehat e_a = \widehat p_{01}$, $\widehat e_n = \widehat p_{10}$, and $\widehat e_c =  \widehat p_{11}-\widehat p_{01}$.

\subsection{Identification formulas}

Although the data generating process involves multiple aspects, point identification of PSCE does not require the specification of all four models. Theorem \ref{thm1} summarizes three different strategies for point identification.

\begin{theorem}\label{thm1}
(Point identification) Suppose Assumptions \ref{assum:2}--\ref{assum:5} hold, $e_g(\bfX)>0$ for all $g\in\{a,c,n\}$, and $S_{zs}^C(u|\bfX)>0$ for all $(z,s)\in \{0,1\}^{\otimes 2}$, the PSCEs are nonparametrically identified.
\begin{compactitem}
    \item[(i)] Using propensity score, principal score, and censoring survival function, we have
    \begin{align*}
\Delta_{c}(u)&= \E\left[\frac{e_{c}(\bfX)}{p_{11}-p_{01}}\frac{S}{p_{11}(\bfX)}\frac{Z}{\pi_1(\bfX)}\frac{\mathbb{I}(U \geq u)}{\mc S^C_{11}(u|\bfX)}\right]- \E\left[\frac{e_{c}(\bfX)}{p_{11}-p_{01}}\frac{1-S}{p_{00}(\bfX)}\frac{1-Z}{\pi_0(\bfX)}\frac{\mathbb{I}(U \geq u)}{\mc S^C_{00}(u|\bfX)}\right],\\
\Delta_{n}(u)&=\E\left[\frac{1-S}{p_{10}}\frac{Z}{\pi_1(\bfX)}\frac{\mathbb{I}(U \geq u)}{\mc S^C_{10}(u|\bfX)}\right]-\E\left[\frac{e_{n}(\bfX)}{p_{10}}\frac{1-S}{p_{00}(\bfX)}\frac{1-Z}{\pi_0(\bfX)}\frac{\mathbb{I}(U \geq u)}{\mc S^C_{00}(u|\bfX)}\right],\\
\Delta_{a}(u)&=\E\left[\frac{e_{a}(\bfX)}{p_{01}}\frac{S}{p_{11}(\bfX)}\frac{Z}{\pi_1(\bfX)}\frac{\mathbb{I}(U \geq u)}{\mc S^C_{11}(u|\bfX)}\right]-\E\left[\frac{S}{p_{01}}\frac{1-Z}{\pi_0(\bfX)}\frac{\mathbb{I}(U \geq u)}{\mc S^C_{01}(u|\bfX)}\right].
\end{align*}
\item[(ii)] Using prinicpal score and outcome survival function, we have
\begin{align*}
\Delta_{c}(u)&=\E\left[\frac{e_{c}(\bfX)}{p_{11}-p_{01}}\{\mc S_{11}(u|\bfX)-\mc S_{00}(u|\bfX)\}\right],\\
\Delta_{n}(u)&=\E\left[\frac{e_{n}(\bfX)}{p_{10}}\{\mc S_{10}(u|\bfX)-\mc S_{00}(u|\bfX)\}\right],\\
\Delta_{a}(u)&=\E\left[\frac{e_{a}(\bfX)}{p_{01}}\{\mc S_{11}(u|\bfX)-\mc S_{01}(u|\bfX)\}\right].
\end{align*}
\item[(iii)] Using propensity score and outcome survival function, we have
\begin{align*}
\Delta_{c}(u)&=\frac{1}{p_{11}-p_{01}}\E\left[\left(\frac{SZ}{\pi_1(\bfX)}-\frac{S(1-Z)}{\pi_0(\bfX)}\right)\{\mc S_{11}(u|\bfX)-\mc S_{00}(u|\bfX)\}\right],\\
\Delta_{n}(u)&=\frac{1}{p_{10}}\E\left[\left(1-\frac{SZ}{\pi_1(\bfX)}\right)\{\mc S_{10}(u|\bfX)-\mc S_{00}(u|\bfX)\}\right],\\
\Delta_{a}(u)&=\frac{1}{p_{01}}\E\left[\frac{S(1-Z)}{\pi_0(\bfX)}\{\mc S_{11}(u|\bfX)-\mc S_{01}(u|\bfX)\}\right].
\end{align*}
\end{compactitem}
\end{theorem}
Theorem \ref{thm1} extends the identification formulas in \cite{ding2017principal} and \cite{jiang2020multiply} to a right-censored time-to-event outcome. The proof is presented in Web Appendix A.2. Using $\Delta_c(u)$ as an example, we provide an intuitive interpretation for each identification formula in Web Appendix A.3. 

Theorem \ref{thm1} motivates three moment-type estimators, $\wh\Delta_g^1(u)$, $\wh\Delta_g^2(u)$, and $\wh\Delta_g^3(u)$, corresponding to the identification strategy (i), (ii) and (iii), respectively. To obtain these estimators, we can replace the true expectation ``$\mathbb{E}$" in each identification formula with the empirical average ``$\mathbb{P}_n$" operator after substituting the true probabilities, $\pi_z(\bfX)$,  $p_{zs}(\bfX)$, $e_g(\bfX)$, $p_{zs}$, $\mc S_{zs}^C(u|\bfX)$, and $\mc S_{zs}(u|\bfX)$, with their corresponding estimators introduced in Section \ref{sec:3.1}, $\widehat{\pi}_z(\bfX)$,  $\widehat p_{zs}(\bfX)$, $\widehat e_g(\bfX)$, $\widehat p_{zs}$, $\widehat{\mc S}_{zs}^C(u|\bfX)$, and $\widehat{\mc S}_{zs}(u|\bfX)$, respectively. We refer to these estimators as singly robust estimators as they are consistent to PSCEs when all associated models are correctly specified and will be biased otherwise.

\subsection{Multiply robust estimation of principal survival causal effects}\label{sec: tr}

The singly robust estimators, $\wh\Delta_g^1(u)$, $\wh\Delta_g^2(u)$, and $\wh\Delta_g^3(u)$, are consistent under $\mc M_{\pi+e+C}$, $\mc M_{e+T}$, and $\mc M_{\pi+T}$, respectively. Here, we use the symbol ``+" to denote the intersection of models such that $\mc M_{\pi+e+C}$ means $\pi_z(\bm X)$, $p_{zs}(\bm X)$, and $S_{zs}^C(t|\bfX)$ are correctly specified. Below, we propose a multiply robust estimator $\wh\Delta_g^{\text{mr}}(u) = \wh{\mc S}_{1,g}^{\text{mr}}(u) - \wh{\mc S}_{0,g}^{\text{mr}}(u)$ to provide robustness against model misspecifications. We use the superscript `$\text{mr}$' to highlight that this estimator is multiply robust in the sense that it does not require all of $\mc M_{\pi}$, $\mc M_{e}$, $\mc M_C$, and $\mc M_T$ to be correctly specified. {To proceed, we denote $\mathcal M_{np}$ as the nonparametric model over the observed data $\mathcal{\bm O}_i = \{\bfX_i,Z_i,S_i,U_i,\delta_i\}$. Then, for a fixed time $u$, we first derive the efficient influence function (EIF) of $\mathcal S_{z,g}(u)$ under the nonparametric model $\mathcal M_{np}$; the detailed proof is given in Web Appendix B.1.

\begin{theorem}\label{thm:eif}
(Efficient influence function) Under Assumptions \ref{assum:2}--\ref{assum:5} and for any $z\in\{0,1\}$, $g\in\{a,c,n\}$, and a fixed $u$ within the valid support of event time, the EIF of $\mathcal S_{z,g}(u)$ under $\mathcal M_{np}$ is given by
\begin{equation*}
\psi_{z,g}^{\text{eif}}(u;\mathcal{\bm O}) = \left\{\psi_{z,g}^{(1)}(u;\mathcal{\bm O}) - \mathcal S_{z,g}(u) \times \psi_{g}^{(2)}(\mathcal{\bm O})\right\}\Big/{e_g},
\end{equation*}
with \begingroup\makeatletter\def\f@size{10.5}\check@mathfonts
\begin{align*}
\psi_{z,g}^{(1)}(u;\mathcal{\bm O}) = & \left[p_{z^*s^*}(\bfX)-kp_{01}(\bfX)\right]\left\{\frac{\mathbb{I}(S=s,Z=z)}{ p_{zs}(\bfX)\pi_z(\bfX) } H_{zs}(u;\mathcal{\bm O}) + \mc S_{zs}(u|\bfX)\right\}\\
 & +   \mc S_{zs}(u|\bfX)\left\{\frac{\mathbb{I}(Z=z^*)}{\pi_{z^*}(\bfX)}[\mathbb{I}(S=s^*)-p_{z^*s^*}(\bfX)]-k\frac{1-Z}{\pi_0(\bfX)}[S-p_{01}(\bfX)]\right\}, \\
\psi_{g}^{(2)}(\mathcal{\bm O}) = & \frac{\mathbb{I}(Z=z^*)}{\pi_{z^*}(\bfX)}[\mathbb{I}(S=s^*)-p_{z^*s^*}(\bfX)]-k\frac{1-Z}{\pi_0(\bfX)}[S-p_{01}(\bfX)] + p_{z^*s^*}(\bfX)-kp_{01}(\bfX),\\
H_{zs}(u;\mathcal{\bm O}) = & \mc S_{zs}(u|\bfX) \left[ \int_{0}^{U \wedge u} \frac{ \Lambda_{zs}(dr|\bfX)}{S_{zs}(r|\bfX)S_{zs}^C(r|\bfX)}-\frac{\delta\mathbb{I}(U < u)}{\mc S_{zs}(U|\bfX)\mc S_{zs}^C(U|\bfX)}\right],
\end{align*}\endgroup
where $\Lambda_{zs}(u|\bfX) = -\log \left\{\mathcal S_{zs}(u|\bfX)\right\}$ is the cumulative hazard function of the outcome event process, and $\{k,s,z^*,s^*\}=[\{1,z,1,1\},\{0,0,1,0\},\{0,1,0,1\}]$ if $g=\{c,n,a\}$, respectively.
\end{theorem}

\begin{remark}
\emph{(A special case of EIF under perfect compliance) Under perfect compliance, the compliers stratum represents the entire study population, implying that $\mathcal S_{z,c}(u) = \mathbb{P}(T(z)\geq u)$ for both $z\in\{0,1\}$; that is, the counterfactual survival functions in the compliers stratum are equivalent to the treatment-specific counterfactual survival functions. Interestingly, the EIFs for the compliers stratum in Theorem \ref{thm:eif} (i.e., $\psi_{z,c}^{\text{eif}}(u;\mathcal O)$ for $z\in\{0,1\}$) degenerates to EIFs for $\mc S_z(u) :=\mathbb{P}(T(z)\geq u)$ under perfect compliance. We demonstrate this by using $\psi_{1,c}^{\text{eif}}(u;\mathcal O)$ as an example, analogous arguments apply to $\psi_{0,c}^{\text{eif}}(u;\mathcal O)$. Specifically, with perfect compliance, we have (i) $e_c=1$ and $Z=S$ for all units, (ii) $p_{zs}(\bfX) = 1$ and 0 if $s=z$ and $s\neq z$, respectively; (iii) $\mathcal S_{zs}(u|\bfX)=\mathbb{P}(T\geq u|Z=z,\bfX) =: \mathcal S_{z}(u|\bfX)$ if $s=z$; and (iv) $\mathcal S_{zs}^C(u|\bfX)=\mathbb{P}(C\geq u|Z=z,\bfX) =: \mathcal S_{z}^C(u|\bfX)$ if $s=z$. Notice $\mathcal S_{zs}(u|\bfX)$ and $\mathcal S_{zs}^C(u|\bfX)$ are not well defined for $s\neq z$ under perfect compliance. Substituting the equalities (i)--(iv) into the expressions of $\psi_{1,c}^{(1)}(u;\mc O)$ and $\psi_{c}^{(2)}(\mc O)$, we obtain
\begin{align*}
\psi_{1,c}^{(1)}(u;\mathcal{\bm O}) = & \left[p_{11}(\bfX)-p_{01}(\bfX)\right]\left\{\frac{SZ}{ p_{11}(\bfX)\pi_1(\bfX) } H_{11}(u;\mathcal{\bm O}) + \mc S_{11}(u|\bfX)\right\}\\
 & +   \mc S_{11}(u|\bfX)\left\{\frac{Z}{\pi_{1}(\bfX)}[S-p_{11}(\bfX)]-\frac{1-Z}{\pi_0(\bfX)}[S-p_{01}(\bfX)]\right\} \\
 = & \frac{Z \mc S_{1}(u|\bfX)}{ \pi_1(\bfX) }  \left[ \int_{0}^{U \wedge u} \frac{ \Lambda_{1}(dr|\bfX)}{S_{1}(r|\bfX)S_{1}^C(r|\bfX)}-\frac{\delta\mathbb{I}(U < u)}{\mc S_{1}(U|\bfX)\mc S_{1}^C(U|\bfX)}\right] + \mathcal S_1(u|\bfX), \\
 \psi_{c}^{(2)}(\mc O) = & \frac{Z}{\pi_{1}(\bfX)}[S-p_{11}(\bfX)]-\frac{1-Z}{\pi_0(\bfX)}[S-p_{01}(\bfX)] + p_{11}(\bfX)-p_{01}(\bfX) = 1,
\end{align*}
where $\Lambda_{1}(u|\bfX)$ is the cumulative hazard function with respect to $\mc S_{1}(u|\bfX)$. Then, noting $\mathcal S_{1,c}(u) = \mathcal S_{1}(u)$ under perfect compliance, $\psi_{1,c}^{\text{eif}}(u;\mathcal O)$ simplifies to
$$
\frac{Z \mc S_{1}(u|\bfX)}{ \pi_1(\bfX) }  \left[ \int_{0}^{U \wedge u} \frac{ \Lambda_{1}(dr|\bfX)}{S_{1}(r|\bfX)S_{1}^C(r|\bfX)}-\frac{\delta\mathbb{I}(U < u)}{\mc S_{1}(U|\bfX)\mc S_{1}^C(U|\bfX)}\right] + \mathcal S_1(u|\bfX) - \mc S_{1}(u).
$$
This is exactly the EIF of $\mc S_{1}(u)$ studied in \citet{bai2013doubly} and \citet{westling2023inference}.} 
\end{remark}

Since the EIF in Theorem \ref{thm:eif} has mean zero, we can solve $\mathbb{E}\left[\psi_{z,g}^{\text{eif}}(u;\mathcal{\bm O})\right]=0$ with respect to $\mathcal S_{z,g}(u)$ and obtain the EIF-induced identification formula:
\begin{equation}\label{eq:eif_identification_formula}
\mathcal S_{z,g}(u) = \mathbb{E}[\psi_{z,g}^{(1)}(u;\mathcal{\bm O})]\Big/ \mathbb{E}[\psi_{g}^{(2)}(\mathcal{\bm O})].
\end{equation}
Therefore, the proposed multiply robust estimator $\wh{\mc S}_{z,g}^{\text{mr}}(u)$ is constructed by 
\begin{equation*}
\wh{\mc S}_{z,g}^{\text{mr}}(u) = \mathbb{P}_n\left[\widehat \psi_{z,g}^{(1)}(u;\mathcal{\bm O})\right] \Big/ \mathbb{P}_n\left[\widehat \psi_{g}^{(2)}(\mathcal{\bm O})\right],
\end{equation*}
where $\widehat \psi_{z,g}^{(1)}(u;\mathcal{\bm O})$ and $\widehat \psi_{g}^{(2)}(\mathcal{\bm O})$ are obtained by substituting the unknown nuisance functions in $ \psi_{z,g}^{(1)}(u;\mathcal{\bm O})$ and $ \psi_{g}^{(2)}(\mathcal{\bm O})$ by their estimators provided in Section \ref{sec:3.1}. An appealing property of $\widehat{\mc S}_{z,g}^{\text{mr}}(u)$ is that it provides three types of protection against partial model misspecification. As demonstrated by the proposition below, $\widehat{\mc S}_{z,g}^{\text{mr}}(u)$ converges to $\mc S_{z,g}(u)$ if either, $\mc M_{\pi+e+C}$, $\mc M_{\pi+T}$, or $\mc M_{e+T}$ is correctly specified, where the proof is given in Web Appendix B.2.

\begin{proposition}\label{thm:triply_robust}
(Triple robustness) Suppose that assumptions \ref{assum:2}--\ref{assum:5} hold. For all $z\in\{0,1\}$ and $g\in\{c,n,a\}$, $\widehat{\mc S}_{z,g}^{\text{mr}}(u)$ is a triply robust estimator for $\mc S_{z,g}(u)$ in the sense that $\widehat{\mc S}_{z,g}^{\text{mr}}(u)$ is consistent to  $\mc S_{z,g}(u)$ under the union model $\mc M_{\pi+e+C}\cup\mc M_{\pi+T}\cup\mc M_{e+T}$. As a result, $\wh\Delta_g^{\text{mr}}(u)$ is also consistent to $\Delta_g(u)$ under $\mc M_{\pi+e+C}\cup\mc M_{\pi+T}\cup\mc M_{e+T}$ for all $g\in\{c,n,a\}$.  
\end{proposition}

The above triple robustness property is general to accommodate quasi-experiments. To unpack this property, we provide some intuition on why the triple robustness property holds in Web Appendix B.3, by explicitly decomposing the asymptotic bias of the numerator and demonstrator of $\widehat{\mc S}_{1,c}^{\text{mr}}(u)$. This decomposition sheds light on the mixed bias property of our estimator, and consists of the basis for potentially deriving a double machine learning estimator that leverages machine learning tools to learn nuisance functions but still achieves root-$n$ inference for PSCE (Web Appendix B.4). Furthermore, in ADAPTABLE, the property of our estimator may be further simplified. That is, since the treatment assignment is randomized, the true propensity score ($\pi_1(\bfX)=\pi_1$) is known and estimating the known propensity score will at most improve the finite-sample efficiency without affecting bias. Because the working propensity score is never misspecified, $\wh{\mc S}_{z,g}^{\text{mr}}(u)$ and $\wh \Delta_{g}^{\text{mr}}(u)$ become doubly robust (rather than triply robust) estimators, in the sense that they are consistent when either $\mc M_{e+C}$ or $\mc M_{T}$ is correctly specified.  

Finally, standard error and confidence intervals for the multiply robust PSCE estimators can be constructed via nonparametric bootstrap. Specifically, we re-sample the entire dataset and re-estimate $\widehat{\mc S}_{z,g}^{\text{mr}}(u)$ for $B$ iterations, where $B$ is a large number (for example, 500 or more). Then, we re-estimate $\widehat{\mc S}_{z,g}^{\text{mr}}(u)$ within each bootstrap dataset. The empirical variance of $\widehat{\mc S}_{z,g}^{\text{mr}}(u)$ among the the $B$ bootstrap datasets is a valid estimator for $\text{Var}\left(\widehat{\mc S}_{z,g}^{\text{mr}}(u)\right)$. Also, the $\frac{\alpha}{2}\times 100\%$ and $(1-\frac{\alpha}{2})\times 100\%$ percentiles of empirical distribution of $\widehat{\mc S}_{z,g}^{\text{mr}}(u)$ among the $B$ bootstrap datasets can be taken as its $(1-\alpha)\times 100$\% confidence interval. Standard error and interval estimators of $\wh \Delta_{g}^{\text{mr}}(u)$ can be constructed similarly.


\begin{remark}
\emph{(Alternative form of the multiply robust estimator) In an earlier version of this manuscript \citep{cheng2023multiply}, we used a projection approach to derive a multiply robust estimator, denoted by $\widehat{\mc S}_{z,g}^{\text{mr*}}(u)$. To provide a context, \cite{jiang2020multiply} developed a multiply multiply robust estimator of $\mc S_{z,g}(u)$ under the complete data $\mathcal{\bm A}_i=\{\bfX_i,Z_i,S_i,T_i\}$ without censoring, and we generalize their estimator to the observed data $\mathcal{\bm O}_i$ when the outcome is right-censored, leveraging projection strategy in Chapter 9 of \cite{tsiatis2006semiparametric}. This alternative estimator is written as
$$
\widehat{\mc S}_{z,g}^{\text{mr*}}(u) = \mathbb{P}_n\left[\widehat \psi_{z,g}^{(1)*}(u;\mathcal{\bm O})\right] \Big/ \mathbb{P}_n\left[\widehat \psi_{g}^{(2)}(\mathcal{\bm O})\right],
$$
where $\psi_{g}^{(2)}(\mathcal{\bm O})$ is defined in Theorem \ref{thm:eif} and $\psi_{z,g}^{(1)*}(u;\mathcal{\bm O})$ is same to $\psi_{z,g}^{(1)}(u;\mathcal{\bm O})$ in Theorem \ref{thm:eif} but replacing $H_{zs}(u;\mathcal{\bm O})$ by
$$
H_{zs}^*(u;\mathcal{\bm O}) =  \mc S_{zs}(u|\bfX) \left[\frac{\mathbb{I}(U\geq u)}{\mc S_{zs}(u|\bfX)\mc S_{zs}^C(u|\bfX)}\!+\! \int_{0}^{u} \frac{dM_{zs}^C(r|\bfX)}{S_{zs}(r|\bfX)S_{zs}^C(r|\bfX)}-1\right],
$$
where $N^C(t)=I(U \leq t,\delta=0)$ is the censoring counting process, $dM_{zs}^C(t|\bfX) = dN^C(t)-\mathbb{I}(U\geq t)d\Lambda^C_{zs}(t|X)$ is the censoring martingale increment, and $\Lambda^C_{zs}(t|\bfX) = -\log \mc S_{zs}^C(t|\bfX)$ is the censoring cumulative hazard function. We demonstrate in Web Appendix B.5 that $\widehat H_{zs}^*(u;\mathcal{\bm O})$ is algebraically equivalent to $\widehat H_{zs}(u;\mathcal{\bm O})$. Therefore, the two multiply robust estimators $\widehat{\mc S}_{z,g}^{\text{mr*}}(u)=\widehat{\mc S}_{z,g}^{\text{mr}}(u)$, even though they appear in different forms.} 
\end{remark}

}

\section{Simulation study}

{
We conduct simulations to evaluate the finite-sample performance of the proposed estimators. {We first consider the scenario under ignorable treatment assignment to mimic a quasi-experiment with unknown propensity scores.} Throughout the simulation, we consider 500 Monte Carlo replications and for each replication, $n=1,000$ observations are simulated based on the following data generation process. We first generate five baseline covariates $\bfX=(X_1,X_2,X_3,X_4,X_5)^T$, where $X_1\sim \mathrm{Bernoulli}(0.5)$, $X_2\sim N(0,1)$, $X_3\sim N(0,1)$, and $(X_4,X_5)^T=(X_2^2-1,X_3^2-1)^T$. Then, the treatment assignment is generated by  $\mathbb{P}(Z=1|\bfX)=1/(1+e^{-\bm\alpha^T\bfX})$, where $\bm{\alpha}=[0,0,0,0.5,0.4]^T$. 
Next, we generate $S$ by the logistic regression $\mathbb{P}(S=1|Z,X)=\frac{\exp\left(-0.5+Z+\bm\gamma^T\bfX\right)}{1+\exp\left(-0.5+Z+\bm\gamma^T\bfX\right)}$, where $\bm\gamma=[0,0,0,0.5,0.4]^T$. 
Within each observed cell $(Z=z,S=s)\in \{0,1\}^{\otimes 2}$, the true failure outcome 
$T$ is generated by $\mc S_{zs}(t|\bfX)=\exp(-te^{-1+0.5s+\bm\psi_{zs}^T\bfX})$, where $\bm\psi_{00}=(0,0,0.2,0.4,0.5)^T$, $\bm\psi_{01}=(0,0,0,0.4,0.2)^T$, $\bm\psi_{10}=(0,0,0,0.4,-0.3)^T$, and $\bm\psi_{11}=(0,0,0,-0.3,0.2)^T$. The censoring time is generated by $\mc S_{zs}^C(t|\bfX)=\exp(-te^{-2+0.3X_4+0.2X_5})$ for each observed cell $(Z=z,S=s)\in \{0,1\}^{\otimes 2}$. 

We evaluate the performance of the multiply robust estimators under 8 different scenarios depending on correct or incorrect specifications of $\mc M_\pi$, $\mc M_e$, $\mc M_C$, and $\mc M_T$. For correctly specified $\mc M_\pi$, $\mc M_e$, $\mc M_C$, or $\mc M_T$, we incorporate all covariates $\bfX$ into the working models $\pi(\bfX)$, $p_{zs}(\bfX)$, $\mc S_{zs}^C(u|\bfX)$, or $\mc S_{zs}(u|\bfX)$, respectively. Under misspecification of working models, we only adjust for the first 3 covariates, $(X_1,X_2,X_3)^T$. The 8 scenarios are indicated in the first column of Table \ref{tab:sim_s0c}. In Scenarios 1--4, the union model $\mc M_{\pi+e+C}\cup\mc M_{\pi+T}\cup\mc M_{e+T}$ is correctly specified so that the multiply robust estimators are consistent by our theory. In Scenarios 5--8, subsets of working models are misspecified such that the multiply robust estimators are not guaranteed to be consistent according to Proposition \ref{thm:triply_robust}.

\begin{table}[htbp]
\centering
\begin{threeparttable}
\caption{Simulation results of $\widehat{\mc S}_{0,c}^{\text{mr}}(u)$, where $\mc M_e, \mc M_\pi, \mc M_T$ and $\mc M_C$ indicate whether the working model for principal score, propensity score, time-to-event and time-to-censoring is correctly specified (denoted by a `T' label) or misspecified (denoted by a `F' label). }\label{tab:sim_s0c}
\begin{tabular}{cc|cccccc}
\hline
                 &              & \phantom{a}\phantom{a}$u$\phantom{a}\phantom{a} & \phantom{a}$\mc{S}_{0,c}(u)$\phantom{a} & \phantom{a}$\wh{\mc{S}}_{0,c}^{\text{mr}}(u)$\phantom{a} & \phantom{a}Bias \phantom{a}  & \phantom{a}Monte Carlo SD & Coverage\phantom{a} \\ \hline
\multicolumn{2}{c|}{Scenario 1} & 1   & 0.695             & 0.695                   & 0.000  & 0.050          & 0.946         \\
$\mc M_e$              & T            & 2   & 0.517             & 0.518                   & 0.000  & 0.048          & 0.960         \\
$\mc M_\pi$            & T            & 3   & 0.397             & 0.395                   & -0.002 & 0.046          & 0.960         \\
$\mc M_T$              & T            & 4   & 0.309             & 0.306                   & -0.003 & 0.039          & 0.968         \\
$\mc M_C$              & T            & 5   & 0.245             & 0.242                   & -0.003 & 0.036          & 0.964         \\ \hline
\multicolumn{2}{c|}{Scenario 2} & 1   & 0.695             & 0.693                   & -0.002 & 0.048          & 0.960         \\
$\mc M_e$              & T            & 2   & 0.517             & 0.517                   & 0.000  & 0.048          & 0.972         \\
$\mc M_\pi$            & T            & 3   & 0.397             & 0.394                   & -0.002 & 0.046          & 0.970         \\
$\mc M_T$              & F            & 4   & 0.309             & 0.307                   & -0.003 & 0.041          & 0.960         \\
$\mc M_C$              & T            & 5   & 0.245             & 0.243                   & -0.002 & 0.037          & 0.970         \\ \hline
\multicolumn{2}{c|}{Scenario 3} & 1   & 0.695             & 0.693                   & -0.002 & 0.172          & 0.936         \\
$\mc M_e$             & F            & 2   & 0.517             & 0.517                   & -0.001 & 0.133          & 0.946         \\
$\mc M_\pi$            & T            & 3   & 0.397             & 0.394                   & -0.003 & 0.108          & 0.956         \\
$\mc M_T$              & T            & 4   & 0.309             & 0.306                   & -0.004 & 0.084          & 0.966         \\
$\mc M_C$              & F            & 5   & 0.245             & 0.241                   & -0.003 & 0.068          & 0.960         \\ \hline
\multicolumn{2}{c|}{Scenario 4} & 1   & 0.695             & 0.692                   & -0.003 & 0.042          & 0.962         \\
$\mc M_e$              & T            & 2   & 0.517             & 0.517                   & 0.000  & 0.044          & 0.954         \\
$\mc M_\pi$            & F            & 3   & 0.397             & 0.395                   & -0.002 & 0.043          & 0.962         \\
$\mc M_T$              & T            & 4   & 0.309             & 0.308                   & -0.002 & 0.038          & 0.968         \\
$\mc M_C$              & F            & 5   & 0.245             & 0.244                   & -0.001 & 0.035          & 0.976         \\ \hline
\multicolumn{2}{c|}{Scenario 5} & 1   & 0.695             & 0.711                   & 0.016  & 0.040          & 0.918         \\
$\mc M_e$              & F            & 2   & 0.517             & 0.537                   & 0.019  & 0.045          & 0.948         \\
$\mc M_\pi$            & T            & 3   & 0.397             & 0.413                   & 0.016  & 0.046          & 0.950         \\
$\mc M_T$              & F            & 4   & 0.309             & 0.322                   & 0.013  & 0.046          & 0.964         \\
$\mc M_C$              & T            & 5   & 0.245             & 0.256                   & 0.011  & 0.045          & 0.966         \\ \hline
\multicolumn{2}{c|}{Scenario 6} & 1   & 0.695             & 0.742                   & 0.047  & 0.030          & 0.692         \\
$\mc M_e$              & T            & 2   & 0.517             & 0.575                   & 0.058  & 0.035          & 0.696         \\
$\mc M_\pi$            & F            & 3   & 0.397             & 0.452                   & 0.055  & 0.036          & 0.696         \\
$\mc M_T$              & F            & 4   & 0.309             & 0.361                   & 0.051  & 0.034          & 0.736         \\
$\mc M_C$              & F            & 5   & 0.245             & 0.292                   & 0.048  & 0.032          & 0.758         \\ \hline
\multicolumn{2}{c|}{Scenario 7} & 1   & 0.695             & 0.595                   & -0.100 & 0.036          & 0.224         \\
$\mc M_e$              & F            & 2   & 0.517             & 0.438                   & -0.079 & 0.032          & 0.340         \\
$\mc M_\pi$            & F            & 3   & 0.397             & 0.335                   & -0.062 & 0.030          & 0.488         \\
$\mc M_T$              & T            & 4   & 0.309             & 0.262                   & -0.047 & 0.026          & 0.612         \\
$\mc M_C$              & F            & 5   & 0.245             & 0.209                   & -0.036 & 0.025          & 0.714         \\ \hline
\multicolumn{2}{c|}{Scenario 8} & 1   & 0.695             & 0.761                   & 0.066  & 0.024          & 0.246         \\
$\mc M_e$              & F            & 2   & 0.517             & 0.600                   & 0.083  & 0.029          & 0.228         \\
$\mc M_\pi$            & F            & 3   & 0.397             & 0.479                   & 0.082  & 0.029          & 0.234         \\
$\mc M_T$              & F            & 4   & 0.309             & 0.387                   & 0.078  & 0.029          & 0.264         \\
$\mc M_C$              & F            & 5   & 0.245             & 0.318                   & 0.073  & 0.029          & 0.328         \\ 
\hline
\end{tabular}
\begin{tablenotes}
\item Monte Carlo SD: Monte Carlo standard deviation; Coverage: 95\% bootstrap confidence interval coverage rate.
\end{tablenotes}
\end{threeparttable}
\end{table}

Table \ref{tab:sim_s0c} summarizes the simulation results of the multiply robust estimator for estimating the counterfactual survival function among the compliers (evaluated at 5 time points) under $z=0$, $\widehat{\mc S}_{0,c}^{\text{mr}}(u)$.  In the first four scenarios, the multiply robust estimator has minimal bias and the associated interval estimator exhibits close-to-nominal coverage rate, even if certain working models are misspecified. This empirically verifies the properties stated in Proposition \ref{thm:triply_robust}. In Scenarios 5--8, $\widehat{\mc S}_{0,c}^{\text{mr}}(u)$ exhibits some bias with attenuated coverage rate as an increasing number of working models are misspecified. Simulation results for other counterfactural survival functions, including $\wh{\mc S}_{1,c}^{\text{mr}}(u)$, $\wh{\mc S}_{0,n}^{\text{mr}}(u)$, $\wh{\mc S}_{1,n}^{\text{mr}}(u)$, $\wh{\mc S}_{0,a}^{\text{mr}}(u)$, $\wh{\mc S}_{1,a}^{\text{mr}}(u)$, are provided in Web Tables 3--7. All results share similar patterns with these in Table \ref{tab:sim_s0c}. Simulation results for PSCEs, i.e., the difference between $\wh{\mc S}_{1,g}^{\text{mr}}(u)$ and $\wh{\mc S}_{0,g}^{\text{mr}}(u)$ for $g\in\{a,n,c\}$, are also similar but omitted for brevity.

For additional comparisons, we have also studied the performance of the three singly robust estimators for estimating $\mc S_{0,c}(u)$, and results for $\widehat{\mc S}_{0,c}^1(u)$, $\widehat{\mc S}_{0,c}^2(u)$, and $\widehat{\mc S}_{0,c}^3(u)$ are summarized in Web Tables 8--10. Each singly robust estimator has minimal bias only under scenarios when all of working models used for constructing the estimator are correctly specified. For example, $\widehat{\mc S}_{0,c}^1(u)$ exhibits small bias in Scenarios 1 and 2 when $\mc M_{\pi+e+C}$ is correct, but can be substantially biased across all remaining 6 scenarios. The multiply robust estimator consistently outperforms the singly robust estimator in terms of the bias. Even in Scenarios 5--8 when two or more working models are misspecified, the multiply robust estimator does not appear to provide notably larger bias than the singly robust estimators. {Web Figure 1 compares the finite-sample relative efficiency of the multiply robust estimator versus the three singly robust estimators for estimating $\widehat{\mc S}_{0,c}(u)$. The relative efficiency is computed as the ratio of the Monte Carlo variance of each singly robust estimator to that of the multiply robust estimator, with values greater than 1 indicating greater efficiency of the multiply robust estimator. When all working models are correctly specified (Scenario 1), $\wh{\mc S}_{0,c}^{\textit{mr}}(u)$ exhibits comparable efficiency with  $\wh{\mc S}_{0,c}^{1}(u)$ and $\wh{\mc S}_{0,c}^{2}(u)$ but is slightly less efficient than $\wh{\mc S}_{0,c}^{3}(u)$. Specifically, relative to $\wh{\mc S}_{0,c}^{1}(u)$ and $\wh{\mc S}_{0,c}^{2}(u)$, the multiply robust estimator $\wh{\mc S}_{0,c}^{\text{mr}}(u)$ achieves relative efficiencies ranging from 1.00 to 1.08 and from 0.93 to 1.05, respectively, across the five time points. In comparison to $\wh{\mc S}_{0,c}^{3}(u)$, its relative efficiency ranges from 0.88 to 0.95. However, this efficiency pattern varies under model misspecification. In Scenario 2, for example, $\wh{\mc S}_{0,c}^{\text{mr}}(u)$ offers  efficiency advantage over $\wh{\mc S}_{0,c}^{2}(u)$ (with the relative efficiency between 6.87 to 12.69), comparable efficiency with $\wh{\mc S}_{0,c}^{1}(u)$ (with relative efficiency between 0.94 to 1.17), but less efficiency to $\wh{\mc S}_{0,c}^{3}(u)$ (with relative efficiency between 0.53 to 0.74). Similarly, in Scenario 3, $\wh{\mc S}_{0,c}^{\text{mr}}(u)$ offers significant efficiency advantage over $\wh{\mc S}_{0,c}^{1}(u)$ and $\wh{\mc S}_{0,c}^{3}(u)$ (with the relative efficiency ranges between 2.25 to 2.54 and between 2.11 to 2.34, respectively), although it offers less efficiency compared to $\wh{\mc S}_{0,c}^{2}(u)$ (with the relative efficiency ranges between 0.85 to 0.94). In Scenario 4,  $\wh{\mc S}_{0,c}^{\text{mr}}(u)$ offers substantial efficiency advantage over $\wh{\mc S}_{0,c}^{1}(u)$ and $\wh{\mc S}_{0,c}^{2}(u)$ (with the relative efficiency ranges between 1.58 to 2.17 and between 1.37 to 4.59, respectively), and offers slight less efficiency with $\wh{\mc S}_{0,c}^{3}(u)$ (with the relative efficiency ranges between 0.94 to 0.95).
Averaged across all scenarios with all singly robust estimators among all time points, the multiply robust estimator maintains a relative efficiency value at 3.68 (with the 10th and 90th percentiles at 0.81 and 7.49), indicating that its relative efficiency does not substantially deteriorate even under model misspecification. We also compare the root mean squared error (RMSE) of the multiply and singly robust estimators of $\mathcal{S}_{0,c}(u)$ in Web Figure 2, where $\widehat{\mathcal{S}}_{0,c}^{\text{mr}}(u)$ does not exhibit inflated RMSE relative to the singly robust estimators across all eight scenarios.
}

To reflect the design in ADAPTABLE, we further evaluate the performance of the proposed estimators under randomization. The data generation process follows the quasi-experiment setting, expect that $Z$ is now randomly simulated from a Bernoulli distribution with $\mathbb{P}(Z=1)=0.5$. The performances of $\wh{\mc S}_{0,c}^{\text{mr}}(u)$, $\wh{\mc S}_{1,c}^{\text{mr}}(u)$, $\wh{\mc S}_{0,n}^{\text{mr}}(u)$, $\wh{\mc S}_{1,n}^{\text{mr}}(u)$, $\wh{\mc S}_{0,a}^{\text{mr}}(u)$, $\wh{\mc S}_{1,a}^{\text{mr}}(u)$, are presented in Web Tables 11--16, respectively. The pattern of simulation results is similar to that in the observational study setting, but now the multiply robust estimator becomes a doubly robust estimator. This doubly robust estimator exhibits minimal bias with nominal coverage rate in Scenarios 1--4 and 7 when the union model $\mathcal M_{e+C} \cup \mc M_{T}$ is correctly specified. }

\section{Applications to the ADAPTABLE pragmatic clinical trial}

\subsection{Description of data and intention-to-treat analysis}

We evaluate the comparative effectiveness of two aspirin dosing strategies, 325 mg versus 81 mg per day, in the ADAPTABLE trial. The ADAPTABLE trial began patient enrollment in 2016, and included 15,076 participants with established atherosclerotic cardiovascular disease that were randomly assigned in a 1:1 ratio to take the high dose or low dose. The participants were followed up until June 2020, with a median duration of follow-up time at 26 months and an interquartile range from 19 to 35 months. {
Web Table 2 summarizes the baseline characteristics among all $n=15,076$ participants stratified by the treatment assignment, which suggests that all baseline characteristics are well balanced in both treatment arms.}

Similar to the primary report of ADAPTABLE, we consider the primary outcome to be the time to first occurrence of death from any reason and hospitalization from stroke or myocardial infarction. We restrict the maximum follow-up time at $t_{\max}=38$ months after randomization, and censor any outcomes beyond this point. {During follow-up, 2,493 participants either withdrew consent or discontinued aspirin intake; for these individuals, the time of withdrawal or discontinuation was treated as the right censoring time. The outcome occurred in 512 participants in the low-dose arm (event rate 6.8\%) and 469 participants in the high-dose arm (event rate 6.2\%). We perform an ITT analysis to estimate the differences of counterfactural survival functions based on the doubly robust approach developed in \cite{bai2013doubly}, where Cox proportional hazard regression is used to model the arm-specific time-to-censoring and time-to-event process conditional on all baseline covariates. We evaluate the ITT effect on a grid of time points with three months intervals ($t= 3, 6, 9, \dots, 36$ months). As shown in Web Figure 3, no differences have been observed between these two aspirin dosages on the time-to-event outcome due to almost overlapping counterfactual survival functions. In ADAPTABLE, 34.8\% of patients who were assigned to the high-dose arm took the low-dose and 6.0\% of patients who were assigned to the low-dose arm took the high-dose. The ITT analysis, however, does not take treatment noncompliance into consideration, and does not offer insights into potential treatment effect heterogeneity across subpopulations defined by compliance patterns.}   

\subsection{Model specification, balance check and strata characteristics}

We apply the proposed methods to investigate the PSCE estimands defined for compliers, always high-dose and always low-dose takers. 
The four models described in Section \ref{sec:3.1} are used to estimate the treatment assignment process, noncompliance status, the time to censoring, and the time to outcome in the ADAPTABLE trial, adjusting for all baseline characteristics. In particular, the working logistic propensity score model is fitted for adjusting any chance imbalance in baseline covariates, rather than addressing confounding due to assignment.

While the treatment is considered randomized, the noncompliance behavior is not randomized and the principal score models are fitted to address compliance-outcome confounding. To check adequacy of the principal score models, we empirically assess the balance for each baseline covariate across before and after principal score weighting; this step is analogous to balance check for propensity score weighted analysis of observational studies \citep{li2018balancing}. 
Specifically, we extend the strategy in \cite{ding2017principal} by defining the following three balancing metrics to quantify the weighted standardized mean differences (SMD) of a given covariate $X$ across the four observed $(Z,S)$ cell, $(Z=1,S=1)$, $(Z=0,S=1)$, $(Z=1,S=0)$, and $(Z=0,S=0)$: 
\begin{align*}
\text{SMD}_c & = \frac{1}{s_c} \left|\mathbb{P}_n\left[\frac{ZS\mc W_{1,c}(\bm X)X}{\mathbb{P}_n[ZS]} - \frac{(1-Z)(1-S)\mc W_{0,c}(\bm X)X}{\mathbb{P}_n[(1-Z)(1-S)]} \right]\right|, \\
\text{SMD}_n & = \frac{1}{s_n} \left|\mathbb{P}_n\left[\frac{Z(1-S)\mc W_{1,n}(\bm X)X}{\mathbb{P}_n[Z(1-S)]} - \frac{(1-Z)(1-S)\mc W_{0,n}(\bm X)X}{\mathbb{P}_n[(1-Z)(1-S)]} \right]\right|, \\
\text{SMD}_a & = \frac{1}{s_a} \left|\mathbb{P}_n\left[\frac{ZS\mc W_{1,a}(\bm X)X}{\mathbb{P}_n[ZS]} - \frac{(1-Z)S\mc W_{0,a}(\bm X)X}{\mathbb{P}_n[(1-Z)S]} \right]\right|,
\end{align*}
where $\mc W_{z,g}(\bm X)$ for $z\in\{0,1\}$ and $g\in\{c,a,n\}$ are weights, the normalization factor is defined as $s_c=\sqrt{\frac{s_{11}^2+s_{00}^2}{2}}$, $s_n=\sqrt{\frac{s_{10}^2+s_{00}^2}{2}}$, $s_a=\sqrt{\frac{s_{11}^2+s_{01}^2}{2}}$, and $s_{zs}^2$ is the empirical variance of $X$ within all subjects in the $(Z=z,S=s)$ cell.
By definition, $\text{SMD}_c$ quantifies the weighted mean difference of $X$ between the $(Z=1,S=1)$ and $(Z=0,S=0)$ cells, $\text{SMD}_n$ quantifies the weighted mean difference of $X$ between the $(Z=1,S=0)$ and $(Z=0,S=0)$ cells, and  $\text{SMD}_a$ quantifies the weighted mean difference of $X$ between the $(Z=1,S=1)$ and $(Z=0,S=1)$ cells. When $\mc W_{z,g}(\bm X) \equiv 1$, the SMDs measure the systematic difference across different $(Z,S)$-strata, and therefore reflects the amount of confounding due to noncompliance. When $\mc W_{z,g}(\bm X)$ is specified as the corresponding principal score weights, that is, $\mc W_{1,c}(\bm X) = \frac{\wh e_c(\bm X)}{\wh p_{11}(\bm X)}\Big/\frac{\wh e_c}{\wh p_{11}}$, $\mc W_{0,c}(\bm X) = \frac{\wh e_c(\bm X)}{\wh p_{00}(\bm X)}\Big/\frac{\wh e_c}{\wh p_{00}}$, $\mc W_{1,n}(\bm X)=1$, $\mc W_{0,n}(\bm X) = \frac{\wh e_n(\bm X)}{\wh p_{00}(\bm X)}\Big/\frac{\wh e_n}{\wh p_{00}}$, and $\mc W_{1,a}(\bm X) = \frac{\wh e_a(\bm X)}{\wh p_{11}(\bm X)}\Big/\frac{\wh e_a}{\wh p_{11}}$, $\mc W_{0,a}(\bm X)=1$, the SMDs quantifies the extent to which the confounding is addressed by weighting.  
In theory, if the principal score models are correctly specified, the SMDs after the weighting converges to 0, and therefore can be used as a diagnostic check for $\wh e_g(\bm X)$. 

\begin{figure}[h]
\begin{center}
\includegraphics[width=0.99\textwidth]{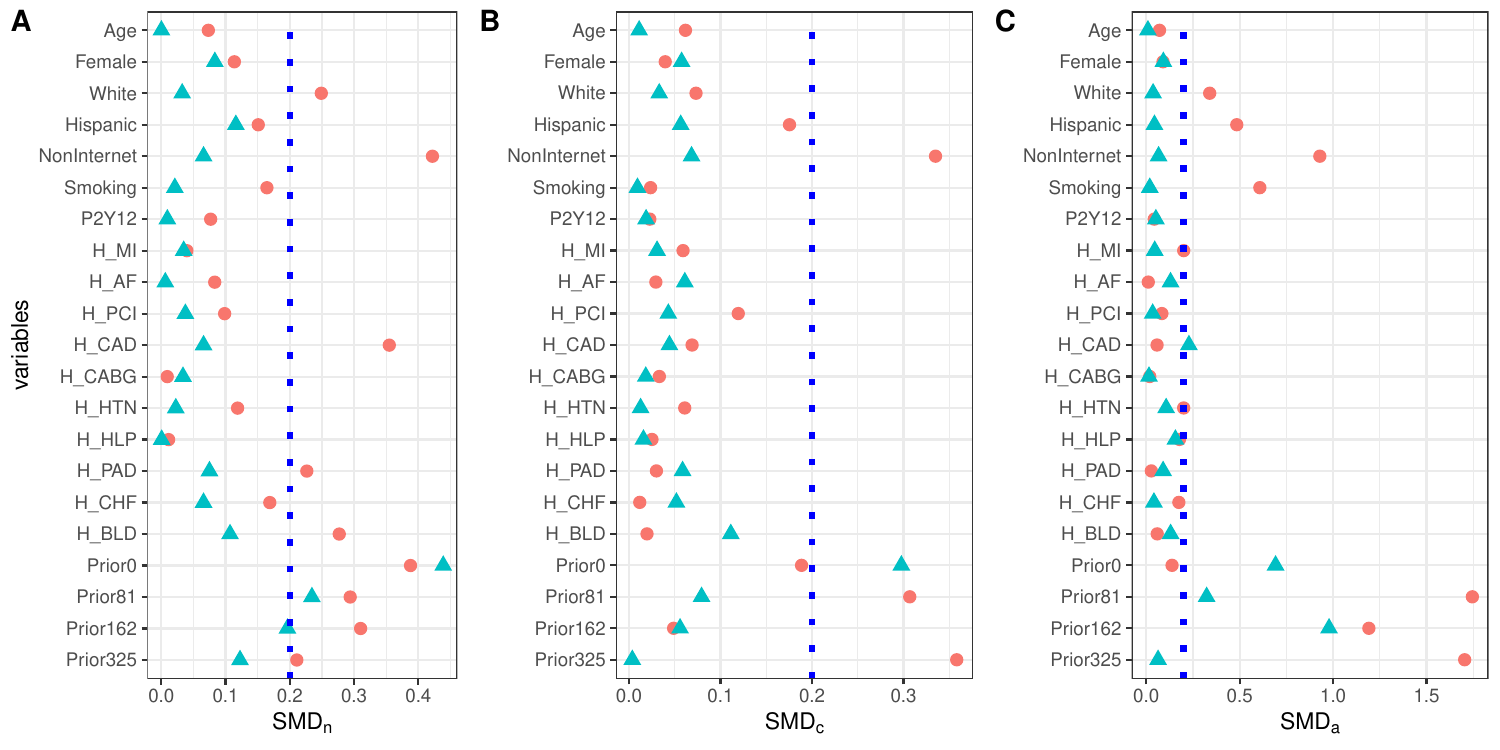}
\caption{Balance check for baseline characteristics for $\text{SMD}_n$ (panel A), $\text{SMD}_c$ (panel B), and $\text{SMD}_a$ (panel C). The red `{\color{WildStrawberry}$\bullet$}' symbol indicates the unweighted SMDs and the blue `{\color{cyan}$\blacktriangle$}' symbol indicates principal score weighted SMDs. The vertical axis shows the names of the variables. Abbreviations: Smoking: current smokers; NonInternet - non-internet users; P2Y12: P2Y12 inhibitor; H\_MI: Myocardial infarction; H\_AF: Atrial fibrillation; H\_PCI: Percutaneous coronary intervention; H\_CAD: Coronary artery disease; H\_CABG: coronary‐artery bypass grafting; H\_HTN: Hypertension; H\_HLP: Hyperlipidemia; H\_PAD: peripheral artery disease; H\_CHF: congestive heart failure; H\_BLD: History of bleeding; Prior0: No prior aspirin intake; Prior81: Prior aspirin dose - 81 mg; Prior162: Prior aspirin dose - 162 mg; Prior325: Prior aspirin dose - 325 mg.  }
\label{fig:balance}
\end{center}
\end{figure}

Figure \ref{fig:balance} provides the love plots of the $\text{SMD}_c$, $\text{SMD}_n$, and $\text{SMD}_a$ for all baseline characteristics before and after principal score weighting. A vertical line is superimposed on this figure to denote a standardized difference of 20\%, which we chose as a threshold for adequate balance. 
{It is evident that the principal score weighting improves the balance of baseline characteristics across the $(Z,S)$-strata. The SMDs after principal score weighting are generally controlled below or around 20\%, with two exceptions on no prior aspirin intake and prior aspirin dosage at 162 mg. 
Specifically, $\text{SMD}_n$ and $\text{SMD}_a$ of no prior aspirin intake are above 0.4, and the $\text{SMD}_a$ of prior aspirin dosage at 162 mg is 0.97. Further inspections indicate that these two are rare covariates (4\% participants have no prior aspirin intake and 2\% participants have a prior dosage at 162mg), and additional attempts on including interactions into the principal score models has limited improvements on the SMD. Therefore we do not further complicate the fitted principal score model.}

\begin{table}[h]
    \centering
    \caption{Mean and standard deviation of baseline characteristics among tha always low-dose takers, compliers, and always high-dose takers. Max ASD is the maximum pairwise absolute standardized difference across the three principal strata. Given a specific baseline covariate, its Max ASD is calculated as the maximum of ${|\bar x_g - \bar x_{g'}|}/{\sqrt{0.5(s_g^2+s_{g'}^2)}}$ for all $g,g'\in \{c,n,a\}$, where $\bar x_g$ and $s_g^2$ are the estimated mean and variance of this covariate in the stratum $g$.}\label{tab:basline_ps}
    \scalebox{1.0}[1.0]{
    \begin{threeparttable}
    \begin{tabular}{lcccc}
    \hline
        Variable & Always low-dose & Compliers & Always high-dose & Max \\ 
        & takers & & takers & ASD\\
        \hline
        Expected sample size & 5231 & 8940 & 905\\
        Age (years) & 67.67 (9.70) & 66.47 (9.85) & 65.58 (10.06) & 0.21 \\
  Female sex & 0.32 (0.47) & 0.31 (0.46) & 0.30 (0.46) & 0.04 \\ 
  White race & 0.76 (0.43) & 0.82 (0.38) & 0.75 (0.43) & 0.17 \\ 
  Hispanic ethnicity & 0.04 (0.19) & 0.03 (0.16) & 0.04 (0.20) & 0.09 \\ 
  Non-internet users & 0.18 (0.39) & 0.09 (0.28) & 0.20 (0.40) & 0.34 \\ 
  Current smokers & 0.10 (0.31) & 0.08 (0.28) & 0.14 (0.35) & 0.18\\
  P2Y12 inhibitor & 0.24 (0.43) & 0.22 (0.41) & 0.23 (0.42) & 0.04 \\ 
  \textbf{Medical History} & & & \\
  ~~Myocardial infarction & 0.38 (0.48) & 0.35 (0.48) & 0.40 (0.49) & 0.09 \\ 
  ~~Atrial fibrillation & 0.09 (0.29) & 0.08 (0.28) & 0.09 (0.29) & 0.04 \\
  ~~Percutaneous coronary intervention & 0.44 (0.50) & 0.39 (0.49) & 0.40 (0.49) & 0.11 \\ 
  ~~Coronary artery disease & 0.95 (0.22) & 0.92 (0.27) & 0.93 (0.25) & 0.11 \\ 
  ~~Coronary‐artery bypass grafting & 0.25 (0.43) & 0.24 (0.43) & 0.24 (0.43) & 0.02 \\ 
  ~~Hypertension & 0.87 (0.34) & 0.84 (0.37) & 0.88 (0.33) & 0.10 \\  
  ~~Hyperlipidemia & 0.88 (0.32) & 0.88 (0.32) & 0.87 (0.34) & 0.06 \\  
  ~~Peripheral artery disease & 0.27 (0.44) & 0.22 (0.42) & 0.23 (0.42) & 0.11 \\ 
  ~~Congestive heart failure & 0.26 (0.44) & 0.23 (0.42) & 0.27 (0.44) & 0.10 \\ 
  ~~History of bleeding & 0.10 (0.30) & 0.08 (0.27) & 0.10 (0.30) & 0.08 \\ 
  \textbf{Aspirin use prior to randomization} & & & \\
  ~~No (prior dose: 0 mg) & 0.03 (0.17) & 0.05 (0.22) & 0.05 (0.21) & 0.11 \\ 
  ~~Prior dose: 81 mg & 0.87 (0.33) & 0.83 (0.38) & 0.44 (0.5) & 1.01 \\ 
  ~~Prior dose: 162 mg & 0.02 (0.13) & 0.02 (0.14) & 0.07 (0.25) & 0.26 \\ 
  ~~Prior dose: 325 mg & 0.08 (0.27) & 0.10 (0.3) & 0.44 (0.5) & 0.90 \\ 
        \hline
    \end{tabular}
    \end{threeparttable}}
\end{table}

Based on the principal score estimates, we calculate the proportion of the three principal strata. {The always low-dose takers and compliers constitute the majority of study population (34.7\% and 59.3\%), and 6.0\% of the study population are always high-dose takers.} To provide further intuitions on each subpopulation, Table \ref{tab:basline_ps} summarizes the moments of baseline characteristic for each principal stratum. The estimated mean of a baseline covariate $X \in \bm X$ in a specific stratum $g\in \{a,c,n\}$ is calculated by $\bar x_{g} = \mathbb{P}_n[\frac{\wh e_g(\bm X)}{\widehat e_g}X]$. Similarly, the estimated standard deviation of $X$ in the stratum $g$ is obtained by the square root of $\mathbb{P}_n[\frac{\wh e_g(\bm X)}{\widehat e_g}(X-\bar x_g)^2]$. {The baseline characteristics are substantially different across principal strata, as quantified by the maximum pairwise absolute standardized difference \citep{li2019propensity} in the last column of Table \ref{tab:basline_ps}. For example, the always low-dose takers are older and include slightly more female patients than the compliers and the always high-dose takers. The compliers include fewer non-internet users and more white patients. The always high-dose takers are the youngest but more smokers. In addition, the always low-dose takers appear to be more likely to have cardiovascular disease histories, with higher prevalence of peripheral artery disease, percutaneous coronary intervention, and coronary artery disease (less healthier subpopulation). Moreover, aspirin use prior to randomization is strongly associated with the aspirin use during the trial and is a strong predictor for the strata membership.}

\subsection{Assessing principal survival causal effects}\label{sec:results}

\begin{figure}[h]
\begin{center}
\includegraphics[width=1\textwidth]{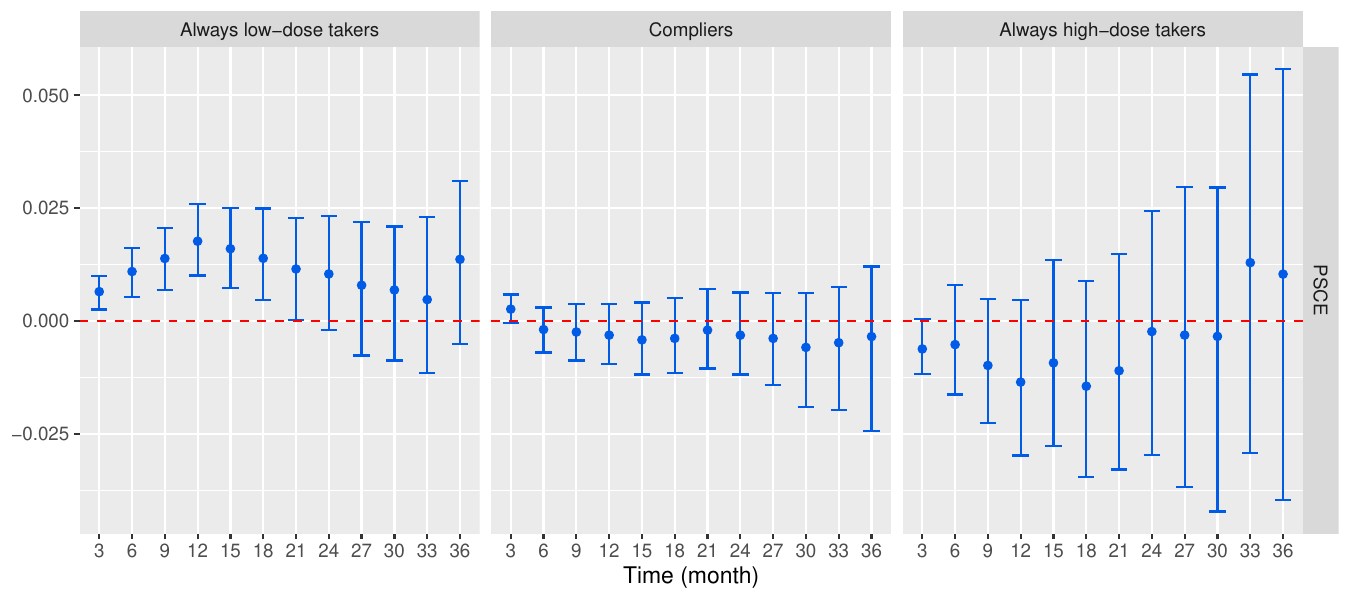}
\caption{
{The principal survival causal effects (PSCEs)  among the always low-dose takers, compliers, and always high-dose takers, evaluated at every three months during follow-up. The expected sample sizes for always low-dose takers, compliers, and always high-dose takers are 5231, 8940, and 905, respectively.} The 95\% confidence bands are based on 500 bootstrap replicates.}
\label{fig:psce_adaptable}
\end{center}
\end{figure}

{We aim to study the PSCEs for a grid of time points with three months intervals until 3 years follow-up ($t= 3, 6, 9, \dots, 36$ months).} Figure \ref{fig:psce_adaptable} presents the PSCE estimates using the proposed multiply robust estimators, where the corresponding counterfactual survival functions within each principal stratum are visualized in Web Figure 4. The 95\% point-wise confidence intervals are obtained by the nonparametric bootstrap with $B=500$ replicates. {The results suggest that the counterfactual survival functions are more separated (compared to ITT) once we stratify by noncompliance patterns. For the always low-dose takers, assignment to the high-dose group leads to a significantly higher survival probability; the magnitude of PSCE peaks at around 12 months ($\widehat\Delta_{n}^{\text{mr}}(12)=0.018$ with 95\% CI from 0.012 to 0.024), after which the effect gradually declines and becomes statistically insignificant for time points at or after 24 months of follow-up. For compliers, assignment to the high-dose group is associated with a small but mostly insignificant harmful effect across time points, with the only exception that a positive PSCE estimate at 3 months has been observed ($\widehat\Delta_{c}^{\text{mr}}(3)=0.002$ with 95\% CI from 0.000 to 0.006). For always high-dose takers, the PSCEs are negative through time points at or before the first 21 months and decline to zero thereafter. However, the confidence intervals for this group include zero at all time points. Notably, the confidence intervals for this stratum are substantially wider than those for the other two strata, likely due to the smaller strata size (only 6.4\% of the study population). In summary, the findings suggest a protective effect of high-dose aspirin assignment among always low-dose takers, potentially reflecting provider-initiated changes in care in response to the assigned dosage. In contrast, the assignment itself does not appear to have a statistically significant impact on mortality or cardiovascular risk among compliers or always high-dose takers.

Under a Bayesian mixture modeling framework and assuming monotonicity, \cite{liu2024principal} also analyzed the PSCEs in the ADAPTABLE trial, which uses a logistic model for principal stratum membership and a Weibull-Cox proportional hazards model for the outcome. Interestingly, despite differences in causal and modeling assumptions, our PSCE findings are qualitatively consistent with theirs. In particular, both analyses suggest positive PSCE estimates for always low-dose takers and small negative PSCE estimates for compliers. For always high-dose takers, \cite{liu2024principal} report strong negative PSCEs throughout the entire follow-up period; our analysis also indicates negative PSCEs before 24 months, but the corresponding confidence intervals include zero.

The above analyses of the PSCEs relies on two assumptions, the principal ignorability and monotonicity, which are not empirically verifiable with observed data alone. Next, we develop sensitivity analysis strategies to evaluate the robustness of our PSCE results when either assumption is violated in the ADAPTABLE trial.
}

\section{Sensitivity analysis for the principal ignorability assumption}

\label{sec:sa_pi}

{We first develop a sensitivity analysis strategy to evaluate robustness of the PSCE estimator under departure from principal ignorability. We consider the following two sensitivity functions to encode the magnitude and direction of departure from principal ignorability:
$$
\ve_1(t,\bfX) = \frac{\mathbb{P}(T(1)\geq t|G=c,\bfX)}{\mathbb{P}(T(1)\geq t|G=a,\bfX)}, \quad \ve_0(t,\bfX) = \frac{\mathbb{P}(T(0)\geq t|G=c,\bfX)}{\mathbb{P}(T(0)\geq t|G=n,\bfX)},
$$
which, in their most general forms, depend on the follow-up time $t$ and covariates $\bfX$.   
Here, $\ve_1(t,\bfX)$ represents the ratio of the potential survival probability under high-dose assignment for compliers to that for always high-dose takers, conditional on the observed covariate level $\bfX$. 
Similar interpretations extend to $\ve_0(t,\bfX)$. Assumption \ref{assum:4} is then equivalent to  $\ve_1(t,\bfX) = \ve_0(t,\bfX) = 1$ for all $t\geq 0$; and $\ve_1(t,\bfX)$ or $\ve_0(t,\bfX) \neq 1$ implies that Assumption \ref{assum:4} does not hold. In Web Appendix C.1, we show that $\mc S_{z,g}(u)$ can be identified based on Assumptions \ref{assum:2}--\ref{assum:3}, \ref{assum:5}, if the sensitivity functions $\{\ve_1(t,\bfX), \ve_0(t,\bfX)\}$ are known. Furthermore, with  known values of $\{\ve_1(t,\bfX), \ve_0(t,\bfX)\}$, we propose a bias-corrected multiply robust estimator $\widehat{\mc S}_{z,g}^{\text{mr-pi}}(u)$ for $\mc S_{z,g}(u)$, and their explicit expressions are summarized in Web Appendix C.1. As suggested by Proposition 2 in Web Appendix C.1, $\widehat{\mc S}_{z,g}^{\text{mr-pi}}(u)$ is doubly robust in the sense that it can provide consistent estimation under either $\mc M_{\pi+e+C}$ or $\mc M_{e+T}$.  

We examine the sensitivity of PSCE estimates in the ADAPTABLE under the violation of the principal ignorability. As $\{\ve_1(t,\bfX), \ve_0(t,\bfX)\}$ are unknown in practice, we consider the following parametric forms of the two sensitivity functions:
$$
\ve_1(t,\bfX) = \exp\left\{\xi_1 \times \frac{t}{t_{\max}}\right\}, \quad \ve_0(t,\bfX) = \exp\left\{\xi_0 \times \frac{t}{t_{\max}}\right\},
$$
where $t_{\max}=38$ months is the maximum follow-up time in ADAPTABLE. More flexible specifications of $\{\ve_1(t,\bfX), \ve_0(t,\bfX)\}$ are given in Web Appendix C.1. Evidently, $\{\xi_1,\xi_0\}$ captures the degree of departure from principal ignorability, with $\xi_1=\xi_0=0$ indicating that principal ignorability holds. Otherwise, principal ignorability is violated when either $\xi_1 \neq 0$ or $\xi_0 \neq 0$, with larger values of $\xi_1$ and $\xi_0$ implying higher degree of violation of  principal ignorability. One can recognize that $\ve_1(t,\bfX)$ is a monotone function that increases (or decreases) from 1 at $t=0$ to $\exp(\xi_1)$ at $t=t_{\max}$, when $\xi_1 > 0$ (or $<0$). Therefore, $\xi_1$ represents the log risk ratio of the outcome at the maximum follow-up time for the compliers versus always high-dose takers. Interpretation of $\ve_0(t,\bfX)$ and $\xi_0$ is analogous. To implement the sensitivity analysis, one can choose grid values of $\{\xi_1,\xi_0\}$ and then summarize $\wh \Delta_{z,g}^{\text{mr-pi}}(u)= \widehat{\mc S}_{1,g}^{\text{mr-pi}}(u)-\widehat{\mc S}_{0,g}^{\text{mr-pi}}(u)$ over the grid of $\{\xi_1,\xi_0\}$ to understand the sensitivity of the PSCE results. Due to low event rate in ADAPTABLE (7.4\% at $t_{\max}=38$ months), if the counterfactual survival function in one stratum is 10\% larger or 10\% lower than its counterpart in another stratum, one of the counterfactual survival function may exceed 1 at $t_{max}$ and hence becomes invalid. Thus, we only allow $\{\xi_1,\xi_0\}$ to vary within the region $[\log(0.9),\log(1.1)]\times [\log(0.9),\log(1.1)]$.

Table \ref{tab:basline_ps} suggests that the always low-dose takers are older and are generally the least healthiest, followed by the the compliers and then the always high-dose takers. If this pattern persists and can be further attributed to unmeasured factors that affect survival, the sensitivity parameters should be restricted within $\xi_1< 0$ and $\xi_0> 0$. Intuitively, this would be the case if unhealthier patients at baseline are concerned about adverse effects that may occur due to taking the high dose, and therefore healthier patients are more willing to adhere to the high-dose assignment. Under this \emph{avoid-harm} scenario, Figure \ref{fig:psce_sa_pi1} presents $\widehat\Delta_{g}^{\text{mr-pi}}(u)$ at four values of $\{\xi_1,\xi_0\}$ within the region $[\log(0.9),0)\times (0,\log(1.1)]$. We observe that the PSCE estimate among always low-dose takers is more robust to such violation of principal ignorability, as $\widehat\Delta_{n}^{\text{mr-pi}}(u)$ is always positive (95\% confidence interval exclude zero) for different values of $\{\xi_1,\xi_0\}$. In contrast, the PSCE estimates among compliers and always high-dose takers are more sensitive to different values of $\{\xi_1,\xi_0\}$, where $\widehat\Delta_{c}^{\text{mr-pi}}(u)$ (and $\widehat\Delta_{a}^{\text{mr-pi}}(u)$) tends to diverge from zero to negative (and positive) values if $\xi_1<0$ and $\xi_0>0$.

\begin{figure}[h]
\begin{center}
\includegraphics[width=0.99\textwidth]{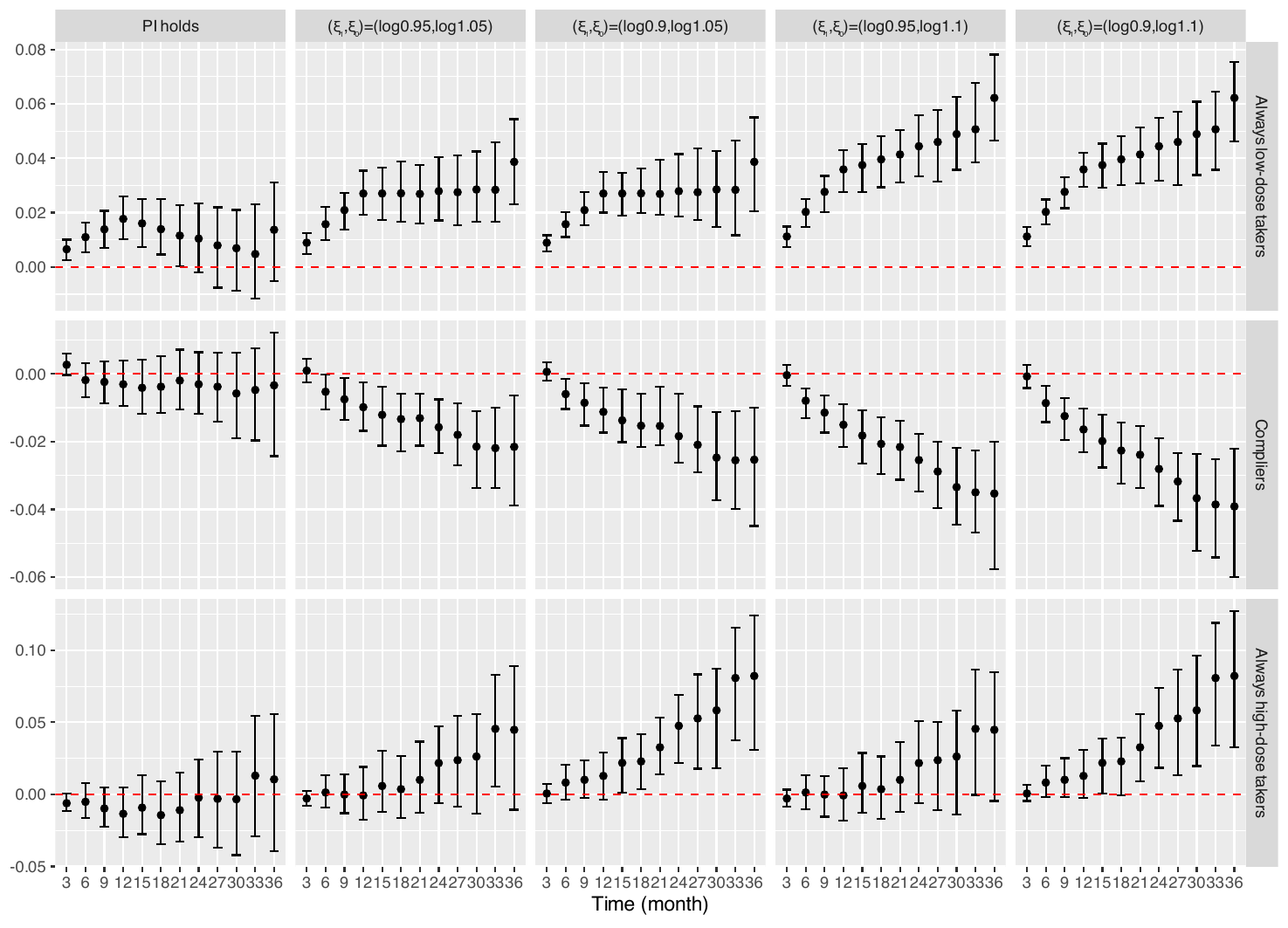}
\caption{Sensitivity of the PSCEs to the violation of the principal ignorability assumption, under the hypothetical scenario that the always low-dose takers are the unhealthiest group and the always high-dose takers are the healthiest group. The PSCEs under principal ignorability assumption are included in the first column for ease of comparison. The expected sample sizes for always low-dose takers, compliers, and always high-dose takers are 5231, 8940, and 905, respectively.}
\label{fig:psce_sa_pi1}
\end{center}
\end{figure}

We further consider an opposite scenario by restricting $\xi_1>0$ and $\xi_0<0$ such that the always low-dose takers are the healthiest and the always high-dose takers are the unhealthiest due to unmeasured factors. Intuitively, this scenario could arise if patients are more concerned about cardiovascular events and, as a result, are willing to tolerate aspirin's side effects because they believe it will help prevent cardiovascular events.
Under this \emph{high-tolerance} scenario, the estimated PSCEs at four values of $\{\xi_1,\xi_0\}$ within the region $(0,\log(1.1)]\times [\log(0.9),0)$ are summarized in Web Figure 5. We observe that the PSCE estimates among all three principal strata are sensitivity to this specific type of violation of principal ignorability. Specifically, $\widehat\Delta_{n}^{\text{mr-pi}}(u)$ for all pre-specified time points attenuate toward the null when $\xi_0=\log(0.95)$ and can move to the opposite direction when $\xi_0=\log(0.9)$. Moreover, $\widehat\Delta_{c}^{\text{mr-pi}}(u)$ (and $\widehat\Delta_{a}^{\text{mr-pi}}(u)$) tends to move from zero to positive (and negative) values when $\xi_1>0$ and $\xi_0<0$.}

\section{Sensitivity analysis for the monotonicity assumption}
\label{sec:SA-Mono}

{If the monotonicity assumption (Assumption \ref{assum:3}) does not hold, we cannot rule out defiers and all four principal strata $G=\{c,n,a,d\}$ need to be considered for estimating PSCEs. We consider the sensitivity function proposed in \citet{ding2017principal} to capture the deviation from monotonicity:
$$
\zeta(\bm X)=\frac{\mathbb{P}(G=d|\bm X)}{\mathbb{P}(G=c|\bm X)},
$$
which takes values from 0 to $\infty$. The sensitivity function $\zeta(\bm X)$ is the ratio between the probabilities of defiers and compliers conditional on $\bm X$. When $\zeta(\bm X)=0$, the monotonicity holds, but $\zeta(\bm X)>0$ implies that the defier strata cannot be ignored. With a fixed value of $\zeta(\bm X)$, we have developed a bias-corrected multiply robust estimator, $\widehat{\mc S}_{z,g}^{\text{mr-mo}}(u)$ for $\mc S_{z,g}(u)$, $z\in\{0,1\}$ and $g=\{c,n,a,d\}$. The  expressions of $\widehat{\mc S}_{z,g}^{\text{mr-mo}}(u)$ are shown in Web Appendix D.1. Similar to Section \ref{sec:sa_pi}, one can report $\wh\Delta_g^{\text{mr-mo}}(u)=\widehat{\mc S}_{1,g}^{\text{mr-mo}}(u)-\widehat{\mc S}_{0,g}^{\text{mr-mo}}(u)$ over a range of values of $\zeta(\bm X)$, which help assess how the PSCE estimates are affected by departures from Assumption \ref{assum:3}. As shown in Proposition 6 in Web Appendix D.1, the bias-correct multiply robust estimator, $\widehat{\mc S}_{z,g}^{\text{mr-mo}}(u)$, remains triply robust under the union model $\mc M_{\pi+e+C}\cup\mc M_{\pi+T}\cup\mc M_{e+T}$.

We investigate the sensitivity of our PSCE analysis in the ADAPTABLE in the presence of defiers. For simplicity, we assume the sensitivity function $\zeta(\bm X)=\zeta$ does not depend on baseline covariates, and choose values of $\zeta$ under the range suggested in Proposition 5 in Web Appendix D.1 to ensure that the proportion of each principal stratum is not negative. Specifically, our previous analysis suggests that $\widehat p_{11} =0.653$ and $\widehat p_{01} = 0.060$, and therefore the range of $\zeta$ are expected to lie between $\left[0,1-\frac{ p_{11}-  p_{01}}{\min(  p_{11},1- p_{01})}\right] \approx [0,0.092]$. This range motivates us to explore $\zeta = \{0,0.05,0.08\}$, and the PSCE estimates becomes unstable if $\zeta>0.08$. 
The results in Web Figure 6 imply that the PSCE estimates among the always low-dose takers and compliers are robust even if there exist defiers. The PSCE among the always high-dose takers, however, are relatively sensitive of $\zeta$. As $\zeta$ increases, $\widehat \Delta_{a}^{\text{mr-mo}}(u)$ is associated with a wider 95\% confidence band. This is likely because the existence of defiers reduces the sample size for the always high-dose takers, increasing the uncertainty for the PSCE estimator among the always high-dose takers.} 

\section{Discussion}

We provide a general approach for identification of the principal causal effects in the presence of treatment noncompliance with a right-censored time-to-event outcome. Instead of using the exclusion restriction, we leverage the principal ignorability assumption and develop a multiply robust approach to identify and estimate the PSCE. 
{We then apply the developed multiply robust estimator to the ADAPTABLE pragmatic clinical trial. Among always low-dose takers, we observe positive PSCE estimates, with 95\% confidence intervals excluding zero at time points prior to 24 months, suggesting a potential protective effect of high-dose aspirin assignment in this subgroup. This effect may partly reflect provider-initiated changes in monitoring or care plans in response to perceived risk under the high-dose assignment, a common feature of open-label pragmatic trials conducted in routine healthcare settings. In contrast, PSCE estimates among compliers and always high-dose takers remain statistically non-significant, as their 95\% confidence intervals include zero throughout follow-up.}

{We recognize that the analysis of ADAPTABLE trial relies on two key identification assumptions. First, the principal ignorability assumption is only appropriate when sufficient baseline covariates have been collected to fully deconfound the relationship between noncompliance and the survival outcome. Based on the assumed sensitivity parameters, we investigate to what extent the estimated PSCE may change under two types of violation of the principal ignorability assumption---the avoid-harm and the high-tolerance scenarios. We found that our PSCE estimate within the always low-dose takers is more robust in the avoid-harm scenario compared to the high-tolerance scenario. In the avoid-harm scenario, the $\Delta_{n}(u)$ estimate is always positive and can even further deviate from the null with larger values of the sensitivity parameter. Nevertheless, the estimated effects are sensitive under the high-tolerance scenarios. Specifically, if unmeasured factors were to confer a 5\% relative increase in survival among compliers compared to always low-dose takers, the observed positive estimates of $\Delta_n(u)$ could be entirely attributed to such confounding and explained away.
Both $\Delta_{a}(u)$ and $\Delta_{c}(u)$ are more sensitive under departure from principal ignorability assumption and tend to deviate from null in both avoid-harm and high-tolerance scenarios. Second, under non-monotonicity, the findings suggest the PSCE estimates within the compliers and always low-dose takers strata are generally robust, but the PSCE estimates among the always high-dose takers are relatively sensitive, possibly due to its attenuated sample size in the presence of defiers.}

{In our sensitivity analysis, we use sensitivity parameters embedded in the sensitivity functions to quantify the degree of departure from key causal assumptions. Specifically, we employ $\zeta$ and $\{\xi_1, \xi_0\}$ to measure violations of monotonicity and principal ignorability, respectively. These parameters are subject to natural bounds. The parameter $\zeta$ must lie within $\left[0, 1 - \frac{{p}_{11} - {p}_{01}}{\min({p}_{11}, 1 - {p}_{01})}\right]$, beyond which the proportion of one principal stratum would be negative. In our ADAPTABLE application, we also found that $\{\xi_1, \xi_0\}$ must lie within $[\log(0.9), \log(1.1)] \times [\log(0.9), \log(1.1)]$, as estimates of the counterfactual survival functions would otherwise exceed 1. As a conservative strategy, we evaluate the sensitivity of PSCE estimates by varying the sensitivity parameters over the entire region defined by these bounds. Additionally, for $\{\xi_1, \xi_0\}$, we restrict attention to the subregions $[\log(0.9), 0) \times (0, \log(1.1)]$ and $(0, \log(1.1)] \times [\log(0.9), 0)$, after reflecting clinical input on the plausible direction of these parameters. 
While the ADAPTABLE analysis demonstrates one approach to operationalizing sensitivity analysis for causal assumptions, it is important to note that sensitivity analysis methods are not unique, and alternative frameworks may yield different insights. For example, one may consider applying benchmarking or calibration framework to derive more precise bounds for the sensitivity parameters \citep{frank2000impact, cinelli2020making}. For example, principal ignorability may be violated in the presence of unmeasured baseline confounders. In such cases, one can consider defining narrower bounds for $\{\xi_1, \xi_0\}$ by benchmarking the strength of unmeasured confounding on the strength of observed baseline covariates in their association with the outcome. Moreover, beyond the proposed sensitivity function approach, other sensitivity analysis approaches can be considered, including the E-value approach \citep{vanderweele2017sensitivity}, Bayesian sensitivity analysis approach \citep{mccandless2007bayesian}, and external validation study approach \citep{sturmer2007adjustments}. Investigation of these alternative methods, especially in the context of principal stratification, represents a valuable direction for future research.}

{Of note, the proposed multiply robust estimator of PSCE incorporates an outcome model ($\mathcal{M}_T$) for the time-to-event process and three other models: a propensity score model ($\mathcal{M}_\pi$) for treatment assignment, a principal score model ($\mathcal{M}_e$) for noncompliance, and a censoring model ($\mathcal{M}_C$) for outcome censoring. Our simulation results demonstrate that no single model dominates the proposed estimator's performance in terms of bias, coverage, or Monte Carlo standard deviation. In scenarios where the estimator is theoretically consistent (Scenarios~1--4), the multiply robust estimator shows minimal bias and nominal coverage. Its finite-sample efficiency also does not substantially decrease under outcome model misspecification (e.g., Scenarios~2, 5, 6, and 8). A modest increase in Monte Carlo standard deviation was observed only in $\widehat{\mathcal S}_{0,c}^{\textit{mr}}(u)$ under Scenario~3 (both $\mathcal{M}_e$ and $\mathcal{M}_C$ are misspecified), but this pattern did not persist across other target estimands. These findings appear to differ from those reported for standard doubly robust estimators of the average treatment effect in settings without censoring, where the estimator is often more sensitive to outcome model misspecification \citep{bang2005doubly}. 
However, the patterns align with previous studies of doubly robust estimators for survival outcomes in the absence of noncompliance; for example, \citet{bai2013doubly} have also demonstrated stable finite-sample performance under outcome model misspecification when estimating survival probability estimands.}

{
In our analysis of ADAPTABLE, we focus on treatment switching as the primary intermediate outcome of interest, treating treatment discontinuation and withdrawal of consent as components of right censoring. Interpreted through the lens of the ICH E9(R1) Estimands Framework \citep{kahan2023eliminating}, our approach corresponds to applying the principal stratum strategy to handle treatment switching, while adopting a hypothetical strategy to address discontinuation and withdrawal as censoring events. We acknowledge, however, that these choices are not unique and that alternative strategies exist for handling post-randomization or intercurrent events. For instance, \citet{liu2024principal} employed the principal stratum strategy to address both treatment switching and discontinuation/withdrawal, using a fully parametric survival mixture model to empirically identify principal causal effects. Their analysis, under additional monotonicity assumptions, characterizes five principal strata: three involving participants who consistently remained on treatment and did not withdraw consent---namely, always low-dose takers, always high-dose takers, and compliers---and two additional strata representing those who discontinued treatment or withdrew consent. Extending our semiparametric efficiency framework to accommodate these five strata would require a fundamental redefinition of the estimands, new identification assumptions, and a restructuring of the efficient influence function. We leave the development of a corresponding multiply robust estimator to future work.

}

{Finally, we focus on point-wise inference for the PSCE at prespecified time points. However, researchers may also be interested in conducting simultaneous inference across an entire time interval. For that purpose, it can be useful to develop uniform confidence bands that allow for valid simultaneous inference of PSCEs across the entire follow-up period. Recent work by \cite{westling2023inference} has proposed strategies to construct uniform confidence bands for the average treatment effects in survival analysis without noncompliance. Extending such techniques to study principal stratification warrants additional work.}

{To facilitate the implementation of the proposed methodology, we have developed the \texttt{mrPStrata} R package along with a concise vignette, which can be accessed at \url{https://github.com/chaochengstat/mrPStrata} and \url{https://rpubs.com/chaocheng/mrPStrata}.}

\section*{Aknowledgement}

A preliminary summary of this work was published in \textit{Clinical Trials} as part of the special issue featuring the conference proceedings of the 15th Annual Conference on Statistical Issues in Clinical Trials (\url{https://doi.org/10.1177/17407745241251773}).

\begin{funding}

Research in this article was supported by a Patient-Centered Outcomes Research Institute Award\textsuperscript{\textregistered} (PCORI\textsuperscript{\textregistered} Award ME-2019C1-16146). The statements presented in this article are solely the responsibility of the authors and do not necessarily represent the official views of PCORI\textsuperscript{\textregistered}, its Board of Governors, or the Methodology Committee. We thank the clinical input and motivating questions from investigators of the ADAPTABLE study team, in particular, Sunil Kripalani, Schuyler Jones, Hillary Mulder. We also thank Yueqi Guo for discussions and contributions during the early stages of this project.
\end{funding}

\begin{supplement}
\stitle{Web Appendix for ``Multiply robust estimation for causal survival analysis with treatment noncompliance''}
\sdescription{The supplementary material is available at \url{https://www.chaochengstat.com}, which contains technical proofs of the results and web tables and figures unshown in the manuscript.}
\end{supplement}


\bibliographystyle{imsart-nameyear} 
\bibliography{refs}       

\begin{thebibliography}{39}

\bibitem[\protect\citeauthoryear{Angelucci and
  Attanasio}{2006}]{angelucci2006estimating}
\begin{btechreport}[author]
\bauthor{\bsnm{Angelucci},~\bfnm{Manuela}\binits{M.}} \AND
  \bauthor{\bsnm{Attanasio},~\bfnm{Orazio~P}\binits{O.~P.}}
(\byear{2006}).
\btitle{Estimating ATT effects with non-experimental data and low compliance}
\btype{Technical Report},
\bpublisher{IZA Discussion Papers}.
\end{btechreport}
\endbibitem

\bibitem[\protect\citeauthoryear{Angelucci, Attanasio and
  Di~Maro}{2012}]{angelucci2012impact}
\begin{barticle}[author]
\bauthor{\bsnm{Angelucci},~\bfnm{Manuela}\binits{M.}},
  \bauthor{\bsnm{Attanasio},~\bfnm{Orazio}\binits{O.}} \AND
  \bauthor{\bsnm{Di~Maro},~\bfnm{Vincenzo}\binits{V.}}
(\byear{2012}).
\btitle{The impact of Oportunidades on consumption, savings and transfers}.
\bjournal{Fiscal Studies}
\bvolume{33}
\bpages{305--334}.
\end{barticle}
\endbibitem

\bibitem[\protect\citeauthoryear{Angrist, Imbens and Rubin}{1996}]{Angrist96}
\begin{barticle}[author]
\bauthor{\bsnm{Angrist},~\bfnm{JD}\binits{J.}},
  \bauthor{\bsnm{Imbens},~\bfnm{GW}\binits{G.}} \AND
  \bauthor{\bsnm{Rubin},~\bfnm{DB}\binits{D.}}
(\byear{1996}).
\btitle{Identification of causal effects using instrumental variables}.
\bjournal{Journal of the American Statistical Association}
\bvolume{91}
\bpages{444--455}.
\end{barticle}
\endbibitem

\bibitem[\protect\citeauthoryear{Bai, Tsiatis and
  O'Brien}{2013}]{bai2013doubly}
\begin{barticle}[author]
\bauthor{\bsnm{Bai},~\bfnm{Xiaofei}\binits{X.}},
  \bauthor{\bsnm{Tsiatis},~\bfnm{Anastasios~A}\binits{A.~A.}} \AND
  \bauthor{\bsnm{O'Brien},~\bfnm{Sean~M}\binits{S.~M.}}
(\byear{2013}).
\btitle{Doubly-Robust estimators of treatment-specific survival distributions
  in observational studies with stratified sampling}.
\bjournal{Biometrics}
\bvolume{69}
\bpages{830--839}.
\end{barticle}
\endbibitem

\bibitem[\protect\citeauthoryear{Baker}{1998}]{baker1998analysis}
\begin{barticle}[author]
\bauthor{\bsnm{Baker},~\bfnm{Stuart~G}\binits{S.~G.}}
(\byear{1998}).
\btitle{Analysis of survival data from a randomized trial with all-or-none
  compliance: estimating the cost-effectiveness of a cancer screening program}.
\bjournal{Journal of the American Statistical Association}
\bvolume{93}
\bpages{929--934}.
\end{barticle}
\endbibitem

\bibitem[\protect\citeauthoryear{Baker and Lindeman}{1994}]{baker1994paired}
\begin{barticle}[author]
\bauthor{\bsnm{Baker},~\bfnm{Stuart~G}\binits{S.~G.}} \AND
  \bauthor{\bsnm{Lindeman},~\bfnm{Karen~S}\binits{K.~S.}}
(\byear{1994}).
\btitle{The paired availability design: a proposal for evaluating epidural
  analgesia during labor}.
\bjournal{Statistics in Medicine}
\bvolume{13}
\bpages{2269--2278}.
\end{barticle}
\endbibitem

\bibitem[\protect\citeauthoryear{Bang and Robins}{2005}]{bang2005doubly}
\begin{barticle}[author]
\bauthor{\bsnm{Bang},~\bfnm{Heejung}\binits{H.}} \AND
  \bauthor{\bsnm{Robins},~\bfnm{James~M}\binits{J.~M.}}
(\byear{2005}).
\btitle{Doubly robust estimation in missing data and causal inference models}.
\bjournal{Biometrics}
\bvolume{61}
\bpages{962--973}.
\end{barticle}
\endbibitem

\bibitem[\protect\citeauthoryear{Breslow}{1972}]{breslow1972discussion}
\begin{barticle}[author]
\bauthor{\bsnm{Breslow},~\bfnm{Norman~E}\binits{N.~E.}}
(\byear{1972}).
\btitle{Discussion of Professor Cox’s paper}.
\bjournal{Journal of Royal Statistical Society: Series B (Statistical
  Methodology)}
\bvolume{34}
\bpages{216--217}.
\end{barticle}
\endbibitem

\bibitem[\protect\citeauthoryear{Cheng et~al.}{2023}]{cheng2023multiply}
\begin{barticle}[author]
\bauthor{\bsnm{Cheng},~\bfnm{Chao}\binits{C.}},
  \bauthor{\bsnm{Guo},~\bfnm{Yueqi}\binits{Y.}},
  \bauthor{\bsnm{Liu},~\bfnm{Bo}\binits{B.}},
  \bauthor{\bsnm{Wruck},~\bfnm{Lisa}\binits{L.}} \AND
  \bauthor{\bsnm{Li},~\bfnm{Fan}\binits{F.}}
(\byear{2023}).
\btitle{Multiply robust estimation for causal survival analysis with treatment
  noncompliance}.
\bjournal{arXiv preprint arXiv:2305.13443v2}.
\end{barticle}
\endbibitem

\bibitem[\protect\citeauthoryear{Cinelli and Hazlett}{2020}]{cinelli2020making}
\begin{barticle}[author]
\bauthor{\bsnm{Cinelli},~\bfnm{Carlos}\binits{C.}} \AND
  \bauthor{\bsnm{Hazlett},~\bfnm{Chad}\binits{C.}}
(\byear{2020}).
\btitle{Making sense of sensitivity: Extending omitted variable bias}.
\bjournal{Journal of the Royal Statistical Society Series B: Statistical
  Methodology}
\bvolume{82}
\bpages{39--67}.
\end{barticle}
\endbibitem

\bibitem[\protect\citeauthoryear{Cuzick et~al.}{2007}]{cuzick2007estimating}
\begin{barticle}[author]
\bauthor{\bsnm{Cuzick},~\bfnm{Jack}\binits{J.}},
  \bauthor{\bsnm{Sasieni},~\bfnm{Peter}\binits{P.}},
  \bauthor{\bsnm{Myles},~\bfnm{Jonathan}\binits{J.}} \AND
  \bauthor{\bsnm{Tyrer},~\bfnm{Jonathan}\binits{J.}}
(\byear{2007}).
\btitle{Estimating the effect of treatment in a proportional hazards model in
  the presence of non-compliance and contamination}.
\bjournal{Journal of the Royal Statistical Society: Series B (Statistical
  Methodology)}
\bvolume{69}
\bpages{565--588}.
\end{barticle}
\endbibitem

\bibitem[\protect\citeauthoryear{Ding and Lu}{2017}]{ding2017principal}
\begin{barticle}[author]
\bauthor{\bsnm{Ding},~\bfnm{Peng}\binits{P.}} \AND
  \bauthor{\bsnm{Lu},~\bfnm{Jiannan}\binits{J.}}
(\byear{2017}).
\btitle{Principal stratification analysis using principal scores}.
\bjournal{Journal of the Royal Statistical Society: Series B (Statistical
  Methodology)}
\bvolume{79}
\bpages{757--777}.
\end{barticle}
\endbibitem

\bibitem[\protect\citeauthoryear{Frangakis and
  Rubin}{1999}]{frangakis1999addressing}
\begin{barticle}[author]
\bauthor{\bsnm{Frangakis},~\bfnm{Constantine~E}\binits{C.~E.}} \AND
  \bauthor{\bsnm{Rubin},~\bfnm{Donald~B}\binits{D.~B.}}
(\byear{1999}).
\btitle{Addressing complications of intention-to-treat analysis in the combined
  presence of all-or-none treatment-noncompliance and subsequent missing
  outcomes}.
\bjournal{Biometrika}
\bvolume{86}
\bpages{365--379}.
\end{barticle}
\endbibitem

\bibitem[\protect\citeauthoryear{Frangakis and Rubin}{2002}]{FrangakisRubin02}
\begin{barticle}[author]
\bauthor{\bsnm{Frangakis},~\bfnm{CE}\binits{C.}} \AND
  \bauthor{\bsnm{Rubin},~\bfnm{DB}\binits{D.}}
(\byear{2002}).
\btitle{Principal Stratification in Causal Inference}.
\bjournal{Biometrics}
\bvolume{58}
\bpages{21--29}.
\end{barticle}
\endbibitem

\bibitem[\protect\citeauthoryear{Frank}{2000}]{frank2000impact}
\begin{barticle}[author]
\bauthor{\bsnm{Frank},~\bfnm{Kenneth~A}\binits{K.~A.}}
(\byear{2000}).
\btitle{Impact of a confounding variable on a regression coefficient}.
\bjournal{Sociological Methods \& Research}
\bvolume{29}
\bpages{147--194}.
\end{barticle}
\endbibitem

\bibitem[\protect\citeauthoryear{Gill, Laan and
  Robins}{1997}]{gill1997coarsening}
\begin{binproceedings}[author]
\bauthor{\bsnm{Gill},~\bfnm{Richard~D}\binits{R.~D.}},
  \bauthor{\bsnm{Laan},~\bfnm{Mark~J}\binits{M.~J.}} \AND
  \bauthor{\bsnm{Robins},~\bfnm{James~M}\binits{J.~M.}}
(\byear{1997}).
\btitle{Coarsening at random: Characterizations, conjectures,
  counter-examples}.
In \bbooktitle{Proceedings of the First Seattle Symposium in Biostatistics}
\bpages{255--294}.
\bpublisher{Springer}.
\end{binproceedings}
\endbibitem

\bibitem[\protect\citeauthoryear{Hirano et~al.}{2000}]{Hirano00}
\begin{barticle}[author]
\bauthor{\bsnm{Hirano},~\bfnm{K}\binits{K.}},
  \bauthor{\bsnm{Imbens},~\bfnm{GW}\binits{G.}},
  \bauthor{\bsnm{Rubin},~\bfnm{DB}\binits{D.}} \AND
  \bauthor{\bsnm{Zhou},~\bfnm{X-H}\binits{X.-H.}}
(\byear{2000}).
\btitle{Assessing the effect of an influenza vaccine in an encouragement
  design}.
\bjournal{Biostatistics}
\bvolume{1}
\bpages{69--88}.
\end{barticle}
\endbibitem

\bibitem[\protect\citeauthoryear{Imbens and
  Angrist}{1994}]{imbens1994identification}
\begin{barticle}[author]
\bauthor{\bsnm{Imbens},~\bfnm{GW}\binits{G.}} \AND
  \bauthor{\bsnm{Angrist},~\bfnm{JD}\binits{J.}}
(\byear{1994}).
\btitle{Identification and estimation of local average treatment effects}.
\bjournal{Econometrica}
\bvolume{62}
\bpages{467--475}.
\end{barticle}
\endbibitem

\bibitem[\protect\citeauthoryear{Jiang, Yang and
  Ding}{2022}]{jiang2020multiply}
\begin{barticle}[author]
\bauthor{\bsnm{Jiang},~\bfnm{Zhichao}\binits{Z.}},
  \bauthor{\bsnm{Yang},~\bfnm{Shu}\binits{S.}} \AND
  \bauthor{\bsnm{Ding},~\bfnm{Peng}\binits{P.}}
(\byear{2022}).
\btitle{Multiply robust estimation of causal effects under principal
  ignorability}.
\bjournal{Journal of the Royal Statistical Society: Series B (Statistical
  Methodology)}
\bvolume{84}
\bpages{1423--1445}.
\end{barticle}
\endbibitem

\bibitem[\protect\citeauthoryear{Jo and Stuart}{2009}]{jo2009use}
\begin{barticle}[author]
\bauthor{\bsnm{Jo},~\bfnm{Booil}\binits{B.}} \AND
  \bauthor{\bsnm{Stuart},~\bfnm{Elizabeth~A}\binits{E.~A.}}
(\byear{2009}).
\btitle{On the use of propensity scores in principal causal effect estimation}.
\bjournal{Statistics in Medicine}
\bvolume{28}
\bpages{2857--2875}.
\end{barticle}
\endbibitem

\bibitem[\protect\citeauthoryear{Jones et~al.}{2021}]{jones2021comparative}
\begin{barticle}[author]
\bauthor{\bsnm{Jones},~\bfnm{W~Schuyler}\binits{W.~S.}},
  \bauthor{\bsnm{Mulder},~\bfnm{Hillary}\binits{H.}},
  \bauthor{\bsnm{Wruck},~\bfnm{Lisa~M}\binits{L.~M.}},
  \bauthor{\bsnm{Pencina},~\bfnm{Michael~J}\binits{M.~J.}},
  \bauthor{\bsnm{Kripalani},~\bfnm{Sunil}\binits{S.}},
  \bauthor{\bsnm{Mu{\~n}oz},~\bfnm{Daniel}\binits{D.}},
  \bauthor{\bsnm{Crenshaw},~\bfnm{David~L}\binits{D.~L.}},
  \bauthor{\bsnm{Effron},~\bfnm{Mark~B}\binits{M.~B.}},
  \bauthor{\bsnm{Re},~\bfnm{Richard~N}\binits{R.~N.}},
  \bauthor{\bsnm{Gupta},~\bfnm{Kamal}\binits{K.}} \betal{et~al.}
(\byear{2021}).
\btitle{Comparative Effectiveness of Aspirin Dosing in Cardiovascular Disease}.
\bjournal{New England Journal of Medicine}
\bvolume{384}
\bpages{1981--1990}.
\end{barticle}
\endbibitem

\bibitem[\protect\citeauthoryear{Kahan et~al.}{2023}]{kahan2023eliminating}
\begin{barticle}[author]
\bauthor{\bsnm{Kahan},~\bfnm{Brennan~C}\binits{B.~C.}},
  \bauthor{\bsnm{Cro},~\bfnm{Suzie}\binits{S.}},
  \bauthor{\bsnm{Li},~\bfnm{Fan}\binits{F.}} \AND
  \bauthor{\bsnm{Harhay},~\bfnm{Michael~O}\binits{M.~O.}}
(\byear{2023}).
\btitle{Eliminating ambiguous treatment effects using estimands}.
\bjournal{American Journal of Epidemiology}
\bvolume{192}
\bpages{987--994}.
\end{barticle}
\endbibitem

\bibitem[\protect\citeauthoryear{Li, Buchanan and
  Cole}{2022}]{li2022generalizing}
\begin{barticle}[author]
\bauthor{\bsnm{Li},~\bfnm{Fan}\binits{F.}},
  \bauthor{\bsnm{Buchanan},~\bfnm{Ashley~L}\binits{A.~L.}} \AND
  \bauthor{\bsnm{Cole},~\bfnm{Stephen~R}\binits{S.~R.}}
(\byear{2022}).
\btitle{Generalizing trial evidence to target populations in non-nested
  designs: Applications to aids clinical trials}.
\bjournal{Journal of the Royal Statistical Society: Series C (Applied
  Statistics)}
\bvolume{71}
\bpages{669}.
\end{barticle}
\endbibitem

\bibitem[\protect\citeauthoryear{Li and Li}{2019}]{li2019propensity}
\begin{barticle}[author]
\bauthor{\bsnm{Li},~\bfnm{Fan}\binits{F.}} \AND
  \bauthor{\bsnm{Li},~\bfnm{Fan}\binits{F.}}
(\byear{2019}).
\btitle{Propensity score weighting for causal inference with multiple
  treatments}.
\bjournal{The Annals of Applied Statistics}
\bvolume{13}
\bpages{2389--2415}.
\end{barticle}
\endbibitem

\bibitem[\protect\citeauthoryear{Li, Morgan and
  Zaslavsky}{2018}]{li2018balancing}
\begin{barticle}[author]
\bauthor{\bsnm{Li},~\bfnm{Fan}\binits{F.}},
  \bauthor{\bsnm{Morgan},~\bfnm{Kari~Lock}\binits{K.~L.}} \AND
  \bauthor{\bsnm{Zaslavsky},~\bfnm{Alan~M}\binits{A.~M.}}
(\byear{2018}).
\btitle{Balancing covariates via propensity score weighting}.
\bjournal{Journal of the American Statistical Association}
\bvolume{113}
\bpages{390--400}.
\end{barticle}
\endbibitem

\bibitem[\protect\citeauthoryear{Liu, Wruck and Li}{2024}]{liu2024principal}
\begin{barticle}[author]
\bauthor{\bsnm{Liu},~\bfnm{Bo}\binits{B.}},
  \bauthor{\bsnm{Wruck},~\bfnm{Lisa}\binits{L.}} \AND
  \bauthor{\bsnm{Li},~\bfnm{Fan}\binits{F.}}
(\byear{2024}).
\btitle{Principal stratification analysis of noncompliance with time-to-event
  outcomes}.
\bjournal{Biometrics}
\bvolume{80}
\bpages{ujad016}.
\end{barticle}
\endbibitem

\bibitem[\protect\citeauthoryear{Loeys and Goetghebeur}{2003}]{loeys2003causal}
\begin{barticle}[author]
\bauthor{\bsnm{Loeys},~\bfnm{Tom}\binits{T.}} \AND
  \bauthor{\bsnm{Goetghebeur},~\bfnm{Els}\binits{E.}}
(\byear{2003}).
\btitle{A causal proportional hazards estimator for the effect of treatment
  actually received in a randomized trial with all-or-nothing compliance}.
\bjournal{Biometrics}
\bvolume{59}
\bpages{100--105}.
\end{barticle}
\endbibitem

\bibitem[\protect\citeauthoryear{McCandless, Gustafson and
  Levy}{2007}]{mccandless2007bayesian}
\begin{barticle}[author]
\bauthor{\bsnm{McCandless},~\bfnm{Lawrence~C}\binits{L.~C.}},
  \bauthor{\bsnm{Gustafson},~\bfnm{Paul}\binits{P.}} \AND
  \bauthor{\bsnm{Levy},~\bfnm{Adrian}\binits{A.}}
(\byear{2007}).
\btitle{Bayesian sensitivity analysis for unmeasured confounding in
  observational studies}.
\bjournal{Statistics in medicine}
\bvolume{26}
\bpages{2331--2347}.
\end{barticle}
\endbibitem

\bibitem[\protect\citeauthoryear{Nguyen, Carlson and
  Stuart}{2024}]{nguyen2024identification}
\begin{barticle}[author]
\bauthor{\bsnm{Nguyen},~\bfnm{Trang~Quynh}\binits{T.~Q.}},
  \bauthor{\bsnm{Carlson},~\bfnm{Michelle~C}\binits{M.~C.}} \AND
  \bauthor{\bsnm{Stuart},~\bfnm{Elizabeth~A}\binits{E.~A.}}
(\byear{2024}).
\btitle{Identification of complier and noncomplier average causal effects in
  the presence of latent missing-at-random (LMAR) outcomes: a unifying view and
  choices of assumptions}.
\bjournal{Biostatistics}
\bvolume{25}
\bpages{978--996}.
\end{barticle}
\endbibitem

\bibitem[\protect\citeauthoryear{Nie, Cheng and Small}{2011}]{nie2011inference}
\begin{barticle}[author]
\bauthor{\bsnm{Nie},~\bfnm{Hui}\binits{H.}},
  \bauthor{\bsnm{Cheng},~\bfnm{Jing}\binits{J.}} \AND
  \bauthor{\bsnm{Small},~\bfnm{Dylan~S}\binits{D.~S.}}
(\byear{2011}).
\btitle{Inference for the effect of treatment on survival probability in
  randomized trials with noncompliance and administrative censoring}.
\bjournal{Biometrics}
\bvolume{67}
\bpages{1397--1405}.
\end{barticle}
\endbibitem

\bibitem[\protect\citeauthoryear{Robins and
  Finkelstein}{2000}]{robins2000correcting}
\begin{barticle}[author]
\bauthor{\bsnm{Robins},~\bfnm{James~M}\binits{J.~M.}} \AND
  \bauthor{\bsnm{Finkelstein},~\bfnm{Dianne~M}\binits{D.~M.}}
(\byear{2000}).
\btitle{Correcting for noncompliance and dependent censoring in an AIDS
  clinical trial with inverse probability of censoring weighted (IPCW) log-rank
  tests}.
\bjournal{Biometrics}
\bvolume{56}
\bpages{779--788}.
\end{barticle}
\endbibitem

\bibitem[\protect\citeauthoryear{St{\"u}rmer
  et~al.}{2007}]{sturmer2007adjustments}
\begin{barticle}[author]
\bauthor{\bsnm{St{\"u}rmer},~\bfnm{Til}\binits{T.}},
  \bauthor{\bsnm{Glynn},~\bfnm{Robert~J}\binits{R.~J.}},
  \bauthor{\bsnm{Rothman},~\bfnm{Kenneth~J}\binits{K.~J.}},
  \bauthor{\bsnm{Avorn},~\bfnm{Jerry}\binits{J.}} \AND
  \bauthor{\bsnm{Schneeweiss},~\bfnm{Sebastian}\binits{S.}}
(\byear{2007}).
\btitle{Adjustments for unmeasured confounders in pharmacoepidemiologic
  database studies using external information}.
\bjournal{Medical care}
\bvolume{45}
\bpages{S158--S165}.
\end{barticle}
\endbibitem

\bibitem[\protect\citeauthoryear{Tong et~al.}{2025}]{tong2025semiparametric}
\begin{barticle}[author]
\bauthor{\bsnm{Tong},~\bfnm{Jiaqi}\binits{J.}},
  \bauthor{\bsnm{Kahan},~\bfnm{Brennan}\binits{B.}},
  \bauthor{\bsnm{Harhay},~\bfnm{Michael~O}\binits{M.~O.}} \AND
  \bauthor{\bsnm{Li},~\bfnm{Fan}\binits{F.}}
(\byear{2025}).
\btitle{Semiparametric principal stratification analysis beyond monotonicity}.
\bjournal{arXiv preprint arXiv:2501.17514}.
\end{barticle}
\endbibitem

\bibitem[\protect\citeauthoryear{Tsiatis}{2006}]{tsiatis2006semiparametric}
\begin{bbook}[author]
\bauthor{\bsnm{Tsiatis},~\bfnm{Anastasios~A}\binits{A.~A.}}
(\byear{2006}).
\btitle{Semiparametric theory and missing data}.
\bpublisher{Springer}.
\end{bbook}
\endbibitem

\bibitem[\protect\citeauthoryear{VanderWeele and
  Ding}{2017}]{vanderweele2017sensitivity}
\begin{barticle}[author]
\bauthor{\bsnm{VanderWeele},~\bfnm{Tyler~J}\binits{T.~J.}} \AND
  \bauthor{\bsnm{Ding},~\bfnm{Peng}\binits{P.}}
(\byear{2017}).
\btitle{Sensitivity analysis in observational research: introducing the
  E-value}.
\bjournal{Annals of internal medicine}
\bvolume{167}
\bpages{268--274}.
\end{barticle}
\endbibitem

\bibitem[\protect\citeauthoryear{Wei et~al.}{2021}]{wei2021estimation}
\begin{barticle}[author]
\bauthor{\bsnm{Wei},~\bfnm{Bo}\binits{B.}},
  \bauthor{\bsnm{Peng},~\bfnm{Limin}\binits{L.}},
  \bauthor{\bsnm{Zhang},~\bfnm{Mei-Jie}\binits{M.-J.}} \AND
  \bauthor{\bsnm{Fine},~\bfnm{Jason~P}\binits{J.~P.}}
(\byear{2021}).
\btitle{Estimation of causal quantile effects with a binary instrumental
  variable and censored data}.
\bjournal{Journal of the Royal Statistical Society: Series B (Statistical
  Methodology)}
\bvolume{83}
\bpages{559--578}.
\end{barticle}
\endbibitem

\bibitem[\protect\citeauthoryear{Westling et~al.}{2023}]{westling2023inference}
\begin{barticle}[author]
\bauthor{\bsnm{Westling},~\bfnm{Ted}\binits{T.}},
  \bauthor{\bsnm{Luedtke},~\bfnm{Alex}\binits{A.}},
  \bauthor{\bsnm{Gilbert},~\bfnm{Peter~B}\binits{P.~B.}} \AND
  \bauthor{\bsnm{Carone},~\bfnm{Marco}\binits{M.}}
(\byear{2023}).
\btitle{Inference for treatment-specific survival curves using machine
  learning}.
\bjournal{Journal of the American Statistical Association}
\bvolume{just-accepted}
\bpages{1--26}.
\end{barticle}
\endbibitem

\bibitem[\protect\citeauthoryear{Yu et~al.}{2015}]{yu2015semiparametric}
\begin{barticle}[author]
\bauthor{\bsnm{Yu},~\bfnm{Wen}\binits{W.}},
  \bauthor{\bsnm{Chen},~\bfnm{Kani}\binits{K.}},
  \bauthor{\bsnm{Sobel},~\bfnm{Michael~E}\binits{M.~E.}} \AND
  \bauthor{\bsnm{Ying},~\bfnm{Zhiliang}\binits{Z.}}
(\byear{2015}).
\btitle{Semiparametric transformation models for causal inference in time to
  event studies with all-or-nothing compliance}.
\bjournal{Journal of the Royal Statistical Society: Series B (Statistical
  Methodology)}
\bvolume{77}
\bpages{397}.
\end{barticle}
\endbibitem

\bibitem[\protect\citeauthoryear{Zeng et~al.}{2021}]{zeng2021propensity}
\begin{barticle}[author]
\bauthor{\bsnm{Zeng},~\bfnm{Shuxi}\binits{S.}},
  \bauthor{\bsnm{Li},~\bfnm{Fan}\binits{F.}},
  \bauthor{\bsnm{Wang},~\bfnm{Rui}\binits{R.}} \AND
  \bauthor{\bsnm{Li},~\bfnm{Fan}\binits{F.}}
(\byear{2021}).
\btitle{Propensity score weighting for covariate adjustment in randomized
  clinical trials}.
\bjournal{Statistics in Medicine}
\bvolume{40}
\bpages{842--858}.
\end{barticle}
\endbibitem

\end{thebibliography}

\end{document}